# Thin Film Charged Particle Trackers

Sungjoon Kim[1], Vikas Berry[1], Jessica Metcalfe[2], and Anirudha V. Sumant[3]

1 Department of Chemical Engineering, University of Illinois at Chicago, 929 W Taylor Street, Chicago, Illinois 60608, USA
2 High Energy Physics Division, 9700 South Cass Avenue, Argonne National Laboratory, Lemont, Illinois 60439, USA
3 Center for Nanoscale Materials, 9700 South Cass Avenue, Argonne National Laboratory, Lemont, Illinois 60439, USA

## Abstract

Silicon tracking detectors have grown to cover larger surface areas up to hundreds of square meters, and are even taking over other sub-detectors, such as calorimeters. However, further improvements in tracking detector performance are more likely to arise from the ability to make a low mass detector comprised of a high ratio of active sensor to dead materials, where dead materials include electrical services, cooling, mechanical supports, etc. In addition, the cost and time to build these detectors is currently large. Therefore, advancements in the fundamental technology of tracking detectors may need to look at a more transformative approach that enables extremely large area coverage with minimal dead material and is easier and faster to build. The advancement of thin film fabrication techniques has the potential to revolutionize the next-to-next generation of particle detector experiments. Some thin film deposition techniques have already been developed and widely used in the industry to make LED screens for TV's and monitors. If large area thin film detectors on the order of several square meters can be fabricated with similar performance as current silicon technologies, they could be used in future particle physics experiments. This paper aims to review the key fundamental performance criteria of existing silicon detectors and past research to use thin films and other semi-conductor materials as particle detectors in order to explore the important considerations and challenges to pursue thin film detectors.

## I. Introduction

Silicon tracking detectors are specialized radiation detectors primarily used in particle physics experiments. A silicon tracker is a device consisting of a sensor and its electronic component, where the penetrating ionizing radiation generates electrical signals in the sensor element, and the interfaced electronics amplify and process the gathered signals. The ionizing radiation generates numerous pairs of charge carriers (electrons and holes) as it passes through the semiconductor materials, which are then swept into the connected electrodes due to the applied electric field, resulting in signal generation. The electronic component then amplifies/stores the electrical signals, which are then used to reconstruct the trajectory of the traveling particle and calculate its momentum. Trackers make up the innermost layer of detectors (figure 1), as their high spatial resolution is most useful in areas with dense particle interactions.[1] As the search goes on for the next new particle in high-energy physics, the need for sensitive, rapid, and durable detectors is increasing. Thin film technology has seen rapid growth from its widespread commercial applications which include transistors, diodes, displays, which are found in almost all modern electronic devices. Compared to conventional silicon trackers, thin film trackers can have advantages such as less cooling requirement, low operating power, and less material consumption in detector fabrication. These advantages make thin film technology an attractive candidate in the fabrication of particle detectors.

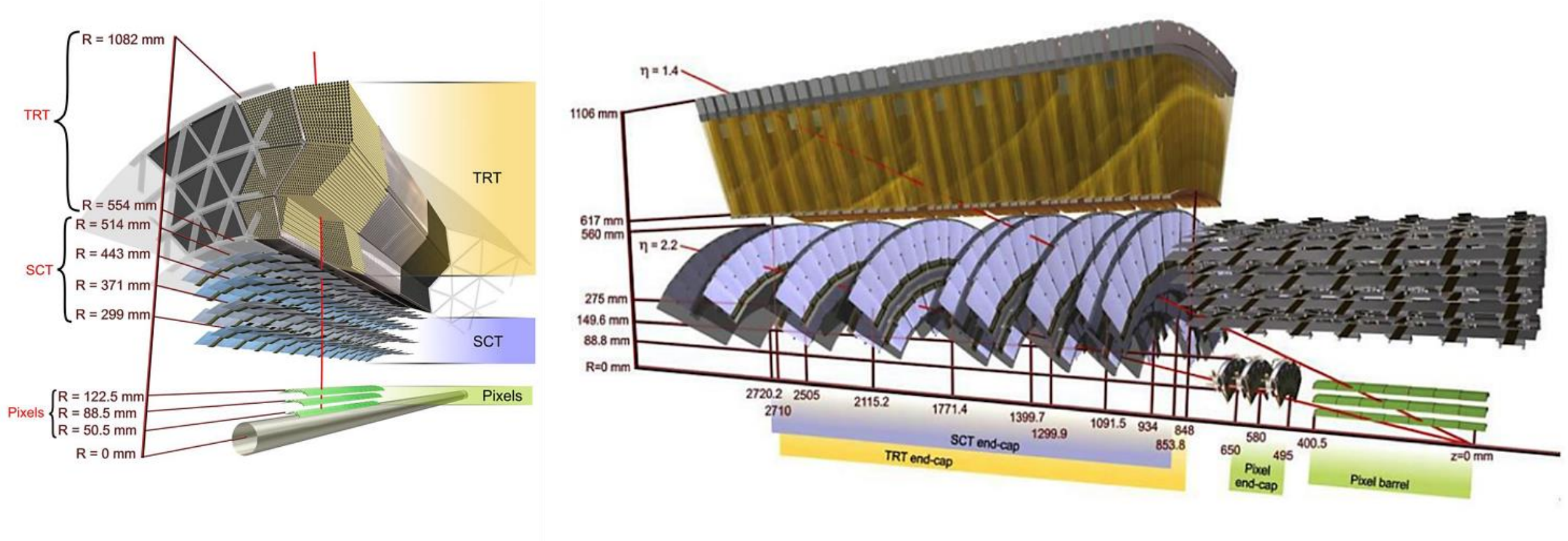

***Figure 1.*** *Schematic view of ATLAS inner detector (ID) barrel (left) and side view of the left end cap (right). Red lines indicate particle tracks. Reproduced with permission from Ref.* [1] *under CC BY 3.0.*

## II. Silicon trackers

Semiconductors have been used for spectroscopic sensor applications since the 1960s.[2] Their detection principle is identical to its predecessor, the gas filled ionization chamber, with the only difference being the detection medium. The primary advantage of changing the detection medium is the increase in energy resolution, as the energy required for producing an electron-hole pair in silicon is 3.6 eV, compared to the 20 eV needed for gas ionization. Towards the late 1970s,[3, 4] semiconductor sensors started seeing more use as tracking sensors in particle physics. For this application, silicon was the material of choice due to its well-studied electrical properties. Some of the other advantages of silicon sensors are (a) it has material properties suitable for particle physics experiments such as low density and radiation length $X_0$. (b) advanced silicon growth techniques are available, enabling the production of large wafers without large impurities/defects which results in large carrier lifetimes (100s of $\mu s$). (c) specific doping profiles of silicon can be achieved, which can be used to optimize the potential shape in the sensor to collect the greatest number of carriers possible. (d) product availability and low price, due to the matured industrial fabrication and processing methods.

Silicon trackers can be categorized into several types, depending on their geometry and the method of connection to their electronic component. The most common and basic geometry is illustrated in Fig.2, where a p+ doped layer is implanted in an n-type bulk. The backside, which has an ohmic contact resulting from a n+ implant on the interfaced metal, is positively biased. This causes a depletion region to form starting from the junction, which expands into the bulk. When an ionizing particle travels through this depletion region, electron-hole pairs are created which are then swept into the electrodes by the electrical field.

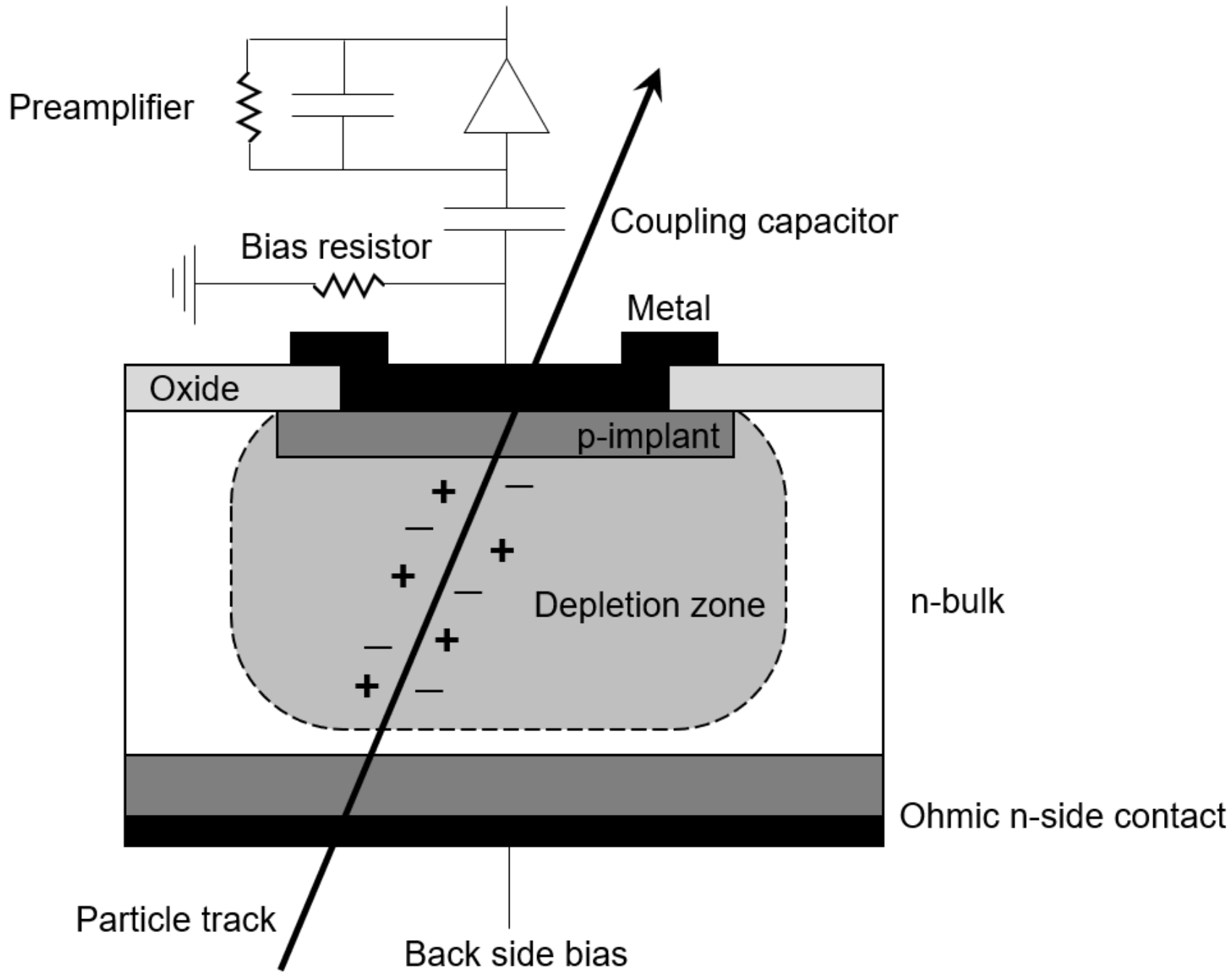

***Figure 2.*** *Cross section of a simplified silicon sensor.*

This type of sensor cannot precisely measure the position of the particle, as it requires one or both electrodes to be segmented. When the electrodes are roughly segmented and the dimension of the element is similar in x and y directions on the surface of the sensor, the device can be classified as a pad detector (figure 2). Pad detectors typically have larger elements compared to pixel detectors and are connected to the electronic component through wire bonds. On the other hand, when each sensor element becomes smaller than a millimeter and the connection to the electronic component is made through bump bonding, these detectors are typically classified as hybrid pixel detectors. If the geometry of the electrodes (width and length) is drastically asymmetric, the device is called a strip detector. A significant difference between strip detectors and pixel detectors is that strip detectors are AC coupled, and the pixel detectors are DC coupled. This is because the AC coupling is not required due to the small leakage currents in each channel.[5]

-Particle types and their effects on the detector

$\beta$ particles at relativistic speeds produce uniform charge clouds when traversing the semiconducting medium, and for this reason is commonly used for laboratory testing of silicon sensors. When particles have particularly low energy or when the absorber is thick, the energy loss of the particle results in the reduction of the particle's speed. This in turn leads to increased ionization (loss of energy), and the cycle repeats. For the case where the particle stops inside the medium, the bulk of its energy is deposited close to the stopping point, creating a localized concentration of electron-hole pairs. This creates a peak in the Bragg curve where the path length of a particle is plotted against stopping power.[6] $\alpha$ particles show shallow penetration, because of their low speed but high charge. These particles typically stop after traveling several micrometers in silicon. The detection of thermal neutrons by conventional methods is difficult, as a) the particle is charge neutral and b) thermal neutrons only interact with the nucleus, leading to a low probability of interaction.[7] However, the incorporation of conversion materials that contain isotopes with large thermal neutron absorption cross sections can enable indirect detection. Some isotopes that satisfy such criteria are boron-10 (3845 barns, alpha), lithium-6 (938 barns, alpha), and gadolinium (49153 barns, beta).[8] Lower atomic numbers are advantageous as they absorb less gamma radiation.[9]

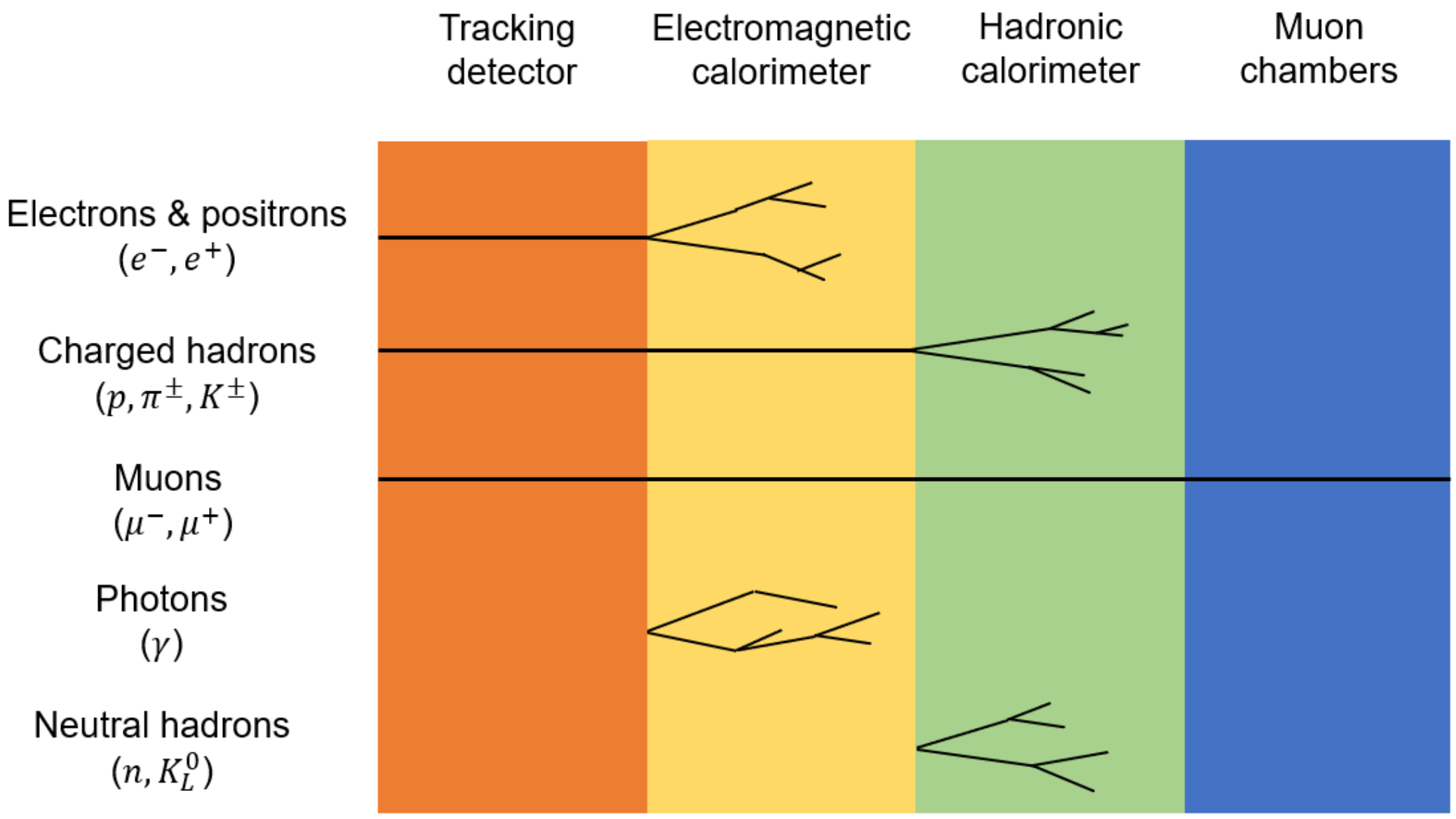

***Figure 3.*** *Typical interaction between particles and detector layers. Tracking detector is the innermost layer, and the Muon chambers form the outermost layer.*

a. Signal generation

- Photons

The energy transfer mechanism from high energy photons (usually X-rays and gamma rays) can be categorized into the following three phenomena, depending on the radiation energy: photoabsorption, Compton scattering, and electron-hole pair production.

For low radiation energy of several hundred KeV, photoabsorption is the dominant energy transfer mechanism. Photoabsorption occurs when the photon transfers all its energy to an electron of the absorbing atom and becomes completely absorbed. The electron is ejected with the excess energy and in the process emits characteristic Auger electrons, or X-rays. This phenomenon becomes much more likely as the atomic number of the absorbing atom becomes higher. When the energy of the photon is between several hundred KeV and ~1 MeV, energy transfer through Compton scattering becomes prevalent. Compton scattering occurs when the photon collides with an electron to transfer some of its energy, resulting in the scattering of the photon as well as the creation of the recoil electron. The likelihood of Compton scattering depends on the material's electron density and increases linearly with its atomic number. When the energy of the photon is above twice the rest mass of electron (over 1.02 MeV), electron-hole pairs may be created. This can occur in the electric field of the electron cloud as well as the nuclei. Until the photon loses enough energy to become photoabsorbed, Compton scattering and electron-hole pair creation continues, resulting in an electromagnetic shower. If the interacting material has an ordered structure (crystalline), electrons in the KeV range may also create electron-hole pairs by exciting the electrons in the atomic shell. If the electron has sufficient energy to ionize the material, free electron and hole carriers are formed. These will then move in the crystal until bound to defects or collected through electrodes to produce electrical signals.

Charged particles
- Energy transfer from particle to sensor medium

The deposition of energy from the charged particle to the sensing material occurs through electron scattering. For particles with energy between mega-electron volts and giga-electron volts, energy loss is well described by the Bethe-Bloch formula, and it is the dominant energy transfer mechanism for particles which are much heavier than electrons.[10, 11]

$$-\langle\frac{dE}{dx}\rangle = Kz^2\frac{Z}{A}\frac{1}{\beta^2}(\frac{1}{2}\ln\frac{2m_ec^2\beta^2\gamma^2T_{max}}{I^2} - \beta^2 + \cdots)$$

where

$\frac{dE}{dx}$ : energy loss of the penetrating particle [$\frac{eV}{g/cm^2}$]

$K$ : $4\pi N_{Av}r_e^2m_ec^2 = 0.307075\ MeV\ cm^2$

$z$ : charge of penetrating particle [electron charge]

$Z$ : atomic number of the sensing medium

$A$ : atomic mass of the sensing medium

$m_ec^2$ : rest mass energy of electron (0.511 MeV)

$\beta$ : particle velocity [speed of light]

$\gamma$ : Lorentz factor ($\frac{1}{\sqrt{1-\beta^2}}$)

$I$ : mean excitation energy (173 eV for silicon)

The large parenthesis denotes the unit for that component. The omitted terms at the end represent correction terms which include the density and shell correction for high and low particle energies, respectively.[10] $T_{max}$ represents the maximum kinetic energy that may be transferred to an electron from a particle with mass M, and is expressed as

$$T_{max} = \frac{2m_ec^2\beta^2\gamma^2}{1 + \frac{2\gamma m_e}{M} + \left(\frac{m_e}{M}\right)^2}$$

When dealing with particles that are much heavier than electrons, the denominator's quadratic term may be ignored. The linear term may also be ignored for most cases, except for high energy particles as in this case $\gamma$ may have similar order as $\frac{m_e}{M}$ and result in large deviations.[5]

The velocity of an ionizing particle affects the amount of energy it deposits on the penetrating material and is plotted through the Bethe-Bloch curve. Very high and low velocities result in increased energy deposition, and the minimum exists between, as a broad valley. Particles with velocities at or near the valley minimum are defined as minimum ionizing particles (MIPs). This is because for particles with higher energies than the MIP, the number of electron hole pairs increase rather slowly (logarithmic) as the energy of the particle is further increased. Thus, most particles with higher energies than the MIP show energy losses similar to the MIP and are treated as such in practice. A MIP for a specific system may be calculated by the following equation

$$MIP = \frac{dE}{dx} \cdot \rho \cdot x, \qquad \frac{MIP}{\omega} = N_{e-h}$$

where $\rho$ and $x$ are the density and thickness of the sensing medium, respectively. MIP for a given sensing medium can be divided by the ionization energy of the medium to calculate the number of electron-hole pairs created ($N_{e-h}$). Ionization energy is the mean energy required for ionization ($\omega$) and is the combination of intrinsic bandgap, optical phonon losses, and residual kinetic energy.[12]

-Mobility-lifetime product ($\mu\tau$)

The drift mobility ($\mu$) of carriers and their lifetimes ($\tau$) are important parameters in radiation detectors because they affect the charge collection efficiency (CCE). The electron-hole pairs generated by the traversing particle experience electrical field from the reverse bias, and are pulled towards cathode and anode, respectively. Whether the charge carriers arrive at the electrodes and result in a signal depends on how fast they can travel through the material, and how long they can survive as separate charge carriers before being destroyed through recombination. Defects in the material structure reduces both the mobility as well as the lifetime.[13] The charge collection distance is defined as:

$$d_{CC} = \mu\tau E$$

where $\mu$, $\tau$, and $E$ are the carrier mobility, carrier lifetime, and applied electric field, respectively. Thus, for radiation detection purposes the carrier mobility-lifetime product is often used to determine the quality of the material (low defect density), as well as consideration points for charge collection distances.

-Scattering

As the particle travels through the detector, it experiences numerous scatterings that lead to small angle deflections. These deflections are mainly due to the Coulomb interaction between the charged particle and the nuclei of the semiconductor medium. The overall scattering angle of the particle after it has interacted with the detector roughly follows a Gaussian distribution[14] with a root mean square of

$$\theta_{plane}^{rms} = \frac{13.6MeV}{\beta pc} \, z \sqrt{\frac{x}{X_0}} [1 + 0.038 \ln(\frac{x}{X_0})]$$

Where $\theta$ is the angle in radian, p is the particle momenta in MeV, $\beta$ is the velocity in units of the speed of light, c. z is the charge number of the ionizing particle, and $\frac{x}{X_0}$ is the semiconductor thickness in units of radiation length. Silicon's radiation length $X_0$ is 9.36 cm.

b. Front-end electronics

Despite the difference in geometries, analog circuits, and readout mechanisms, the currently available pixel chips have several common characteristics. The *active area* consists of matrices of rectangular/square pixels, and the *chip periphery* is responsible for the control of the active area, namely collecting data and housing the global functions for the pixels. Wire bonding pads are located at one edge, so that other chips may be positioned side by side on the module.

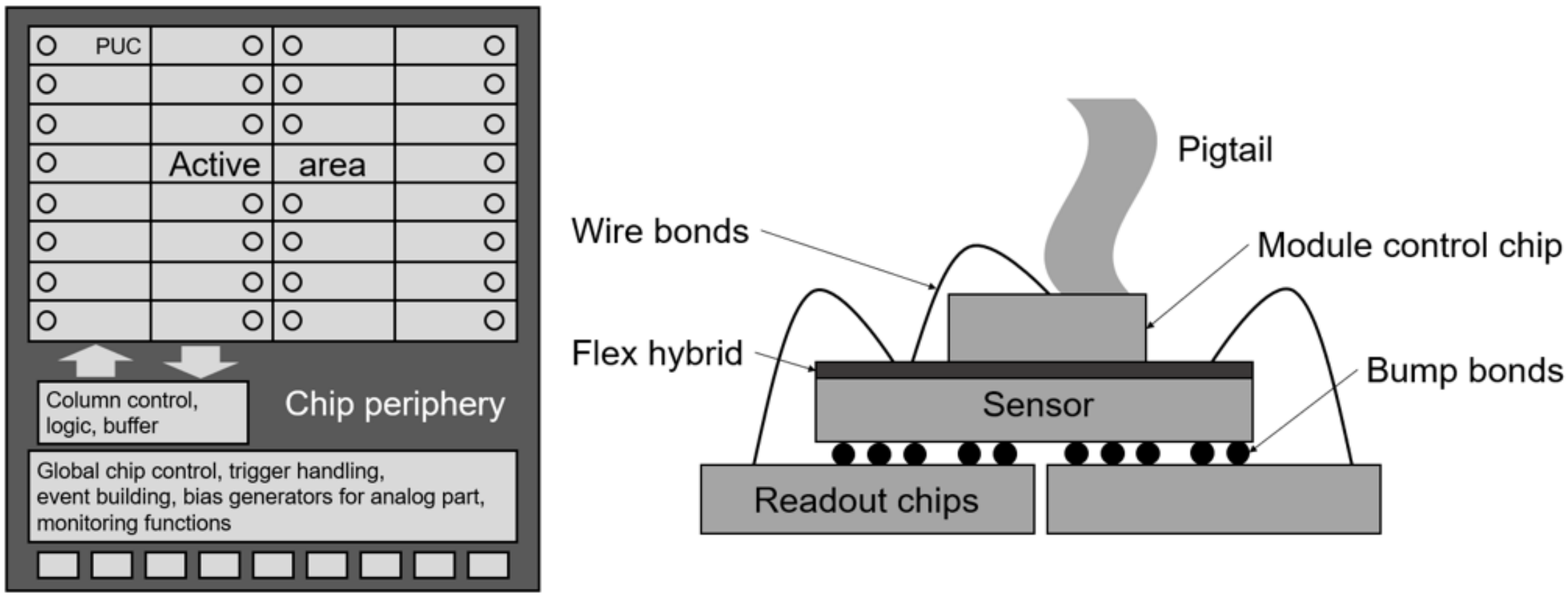

***Figure 4.*** *Geometry of a pixel chip – arrays of identical ships populate the surface of a module (Left) and a cross section of a hybrid module(Right).*

Pixel unit cells (PUCs) which populate the active area of the chip typically have the same dimension as the corresponding pixel sensor. Most signals (power, bias, etc.) and data flows are routed vertically from the grouped PUCs, which are usually grouped in columns. This approach is useful because it minimizes the area taken up by bus signals in the PUCs. A pair of columns can be grouped together to share common circuitry between the pixels while also reducing the cross talk that can occur between the analog and digital areas. This columnar layout with its signal components located at the bottom of the chip allows the edge of the chip to be cut as close as the fabrication technology allows (which is approximately 50 $\mu m$), which in turn enables the placement of as many chips as possible on a single module.

The bottom part of the chip, or the *chip periphery*, can be divided into three main sections. The repeating blocks that are responsible for interfacing the columns above, a section that governs the global control and bias, and finally the wire bond pads. The interfacing section distributes bias signals to the columns and PUCs. It also sends buffered control signals and sometimes contain buffer memory to store the data sent from the pixels. The global control is the communication channel between the chip and the world. A serial protocol is almost always used to reduce the number of wire bond pads. It can provide configuration data to the chip and relay the hit information from the chip to the data acquisition (DAQ).

c. Sensor-electronic connection

i. Bump bonding

The readout chips are connected to the sensors through connections called bump-bonds, and the resulting devices are called hybrid pixel detectors. Bump bonding is a technique used commonly in the electronic industry to connect integrated circuit dies to PCBs, and offers high connection density and small minimum pitch.[15] The process of bump bonding occurs through the following processes: (1) metal droplet deposition onto the pads of electronics. (2) alignment of the electronics chip with the sensor, such that the pads of the pixel face the pads of the electronics. (3) application of pressure and heat to establish permanent electrical connection. For particle physics applications, solder and indium bump bonding are most common. Solder bump bonding requires multiple steps and a higher process temperature[16, 17], whereas indium bump bonding is useful for low temperature connections while suffering from higher costs and lower yields.[18] For small-scale experimental setups, gold stud bump bonding is a viable option due to its inert nature as well as malleability.[19] Soft, inhomogeneous and exotic substrates benefit from this process. Hybrid pixel technology that uses bump bonding has been the choice of detectors for high energy physics applications. Hybrid pixel detectors several advantages which include: (1) better physics performance due to lower mass (2) streamlined production process (3) availability from various vendors due to being a developed, industrial process. (Lower cost) (4) the ability to test the devices during assembly, leading to higher device yields.

ii. Monolithic sensors

For the future upgrades to the large hadron collider (LHC), pixel detectors will have larger area coverage than now. Achieving this with pure hybrid pixel technology poses a significant material issue (sensor, readout chips, module flex), as well as an economic cost issue for larger areas. For silicon pixel detectors, the sensor and the readout electronic circuitry share the same material and thus it is natural to consider the development of a monolithic pixel detector, which combines the production of sensing and reading components. Such design has several important advantages over the hybrid pixel design, including low input capacitances leading to low noise, lower production costs due to the elimination of the complex/costly bonding (hybridization) process. The idea of a monolithic sensor has been considered since the 1980s[20–24], but developments in sensors for LHC-type applications have been mostly restricted to laboratory test-scale chips. Several challenges remain before the realization of monolithic pixels. For hybrid pixel technology, the main limitations are from the electronic circuitry for power density, and the bumping technology for the pitch. For a 0.1 $\mu m$ CMOS technology and a square geometry, the projected pixel size can reach $10 \times 80\ \mu m^2$ or $25 \times 25\ \mu m^2$.[5] This results in an approximate power density of 30 $kW/m^2$. The pitch limit for PbSn solder bumping is in the range of $10 \sim 15\ \mu m$, due to the precision of galvanic process as well as the mask alignment. The pitch limit for indium bumps is slightly lower.[25] Hybrid pixel technology has access to high-density electronic circuitry, where hundreds of transistors are connected to each pixel. For monolithic devices, the number of transistors that can be incorporated is currently much less, restricting the complexity of logic on the device. Despite its current limitations, future detector developments will likely occur through monolithic type detectors, and different thin film materials will be incorporated into the fabrication process.

## III. Thin film detectors

The idea of incorporating different materials into particle trackers is not new, and various scales of efforts have been made to investigate the feasibility of several materials other than silicon. This section will summarize some of the key works done with each detector material. Table 1 lists the properties of some of the potential detector materials, both elemental and compound.

*Table 1. Potential charged particle detector materials and their properties.*[26–29]

| Material | Density ($g/cm^3$) | Band gap (eV) | Intrinsic carrier concentration ($cm^{-3}$) | Average atomic number | Ionization energy (eV) | Drift Mobility ($cm^2/(Vs)$) | | Carrier lifetime | MIP in 10 $\mu m$ (keV) |
|---|---|---|---|---|---|---|---|---|---|
| | | | | | | Electron | Hole | | |
| As | 5.73 | | | 33 | | | | | 7.85 |
| B | 2.37 | | | 5 | | | | | 3.85 |
| Diamond | 3.51 | 5.48 | $< 10^3$ | 6 | 13.1 | 1,800 | 1,200 | ~ 1 ns | 6.25 |
| Cd | 8.65 | | | 48 | | | | | 11.05 |
| Ga | 5.90 | | | 31 | | | | | 8.14 |
| Ge | 5.32 | 0.66 | $2.4 \times 10^{13}$ | 32 | 2.96 | 3,900 | 1,800 | 250 $\mu s$ | 7.29 |
| I | 4.93 | | | 53 | | | | | 6.23 |
| Pb | 11.35 | | | 82 | | | | | 12.73 |
| S | 2.00 | | | 16 | 6.64* | | | | 3.30 |
| a-Se | 4.3 | 2.3 | | 34 | 7 | 0.005 | 0.14 | 10 $\mu s$ | |
| Si | 2.33 | 1.12 | $1.45 \times 10^{10}$ | 14 | 3.61 | 1,415 | 480 | ~ 250 $\mu s$ | 3.9 |
| a-Si | 2.15 | 1.5 ~ 1.8 | | 14 | 4.8 ~ 6 | 1 ~ 5 | 0.01 | ~ $\mu s$ | 3.6 |
| Zn | 7.13 | | | 30 | 8.1* | | | | 10.06 |
| | | | | | | | | | |
| CdTe | 6.1 | 1.44 | $10^7$ | 50 | 4.43 | 1,050 | 100 | 0.1-2 $\mu s$ | 7.81 |
| CdS | 4.8 | 2.42 | | 32 | 6.3 | 340 | 50 | | 19.08 |
| CdSe | 5.81 | 1.73 | | 41 | 5.5* | 720 | 75 | ~ $\mu s$ | |
| CdZnTe | 6 | ~ 1.6 | $10^7$ | 43.3 | 4.6 | ~1,000 | 50-80 | ~ $\mu s$ | 29.8 |
| CdZnSe | 5.5 | 1.7 ~ 2.7 | | 37.3 | | | | | |
| GaAs | 5.4 | 1.42 | $1.8 \times 10^6$ | 32 | 4.35 | 8,800 | 320 | 1-10 ns | 7.45 |
| GaSe | 4.55 | 2.03 | | 32.5 | 4.5 | 75 | 45 | 35 $\mu s$ | |
| HgBrI | 6.2 | 2.4 ~ 3.4 | | 56 | | | | | |
| $HgI_2$ | 6.3 | 2.13 | | 66.5 | 4.2 | 100 | 4 | ~ $\mu s$ | 35.8 |
| HgTe | 8.1 | 0 | | 66 | | 22,000 | 100 | | 54.7 |
| InAs | 5.7 | 0.36 | | 41 | 1.94* | 33,000 | 460 | | 26.8 |
| IGZO | 6 | | | 29.5 | 7.58* | 15 | 0.1 | | |

*Table 2. Potential charged particle detector materials and their properties.*[26–29] *(Continued)*

| Material | Density ($g/cm^3$) | Band gap (eV) | Intrinsic carrier concentration ($cm^{-3}$) | Average atomic number | Ionization energy (eV) | Drift Mobility ($cm^2/(Vs)$) | | Carrier lifetime | MIP in 10 $\mu m$ (keV) |
|---|---|---|---|---|---|---|---|---|---|
| | | | | | | Electron | Hole | | |
| InI | 5.31 | 2.01 | | 51 | | 16600 | | 6 ns | |
| InP | 4.8 | 1.35 | $1.3 \times 10^7$ | 32 | 4.2 | 4,600 | 150 | | 20.5 |
| InSb | 5.8 | 0.17 | | 50 | 1.57* | 78,000 | 750 | | 28.1 |
| PbS | 7.6 | 0.41 | | 49 | 1.98* | 6,000 | 4,000 | | 46.8 |
| $PbI_2$ | 6.2 | 2.32 | | 67.5 | 4.9 | 8 | 2 | 8 $\mu s$ | |
| TlBr | 7.56 | 2.68 | | 58 | 6.5 | 6 | | 2.5 $\mu s$ | |
| TlBrI | 7.5 | 2.2 ~ 2.8 | | 56.3 | | ~ 4.5 | | ~ 2 $\mu s$ | |
| ZnO | 5.6 | 3.37 | | 19 | 8.25* | 130 | - | | 24.8 |
| ZnS | 4.1 | 3.68 | | 23 | 8.23 | 165 | 5 | | |
| ZnTe | 5.72 | 2.26 | | 41 | 7.0* | 340 | 100 | 4 ns | |

*Estimated values

a. Semiconducting materials

i. Hydrogenated amorphous Silicon (a-Si:H) and silicon on insulator

For radiation and particle detection, crystalline silicon currently represents the gold standard. The hydrogenated amorphous silicon (a-Si:H) is an alternative silicon material which has seen vast development in recent decades, resulting in transistors, photovoltaic cells, and memory devices. [30–32] Incorporation of a-Si:H into particle detectors has been explored since the mid ‘80s in the form of p-i-n (p-doped-intrinsic doping-n-doped) structures. Plasma enhanced chemical vapor deposition (PECVD) allows wafer-scale (> 400 mm diameter) deposition of a-Si:H at relatively low temperatures of between 150 and 250 °C, which makes it possible to directly deposit material onto various substrates including glass, metal foils, polymers, as well as existing application specific integrated circuits (ASICs). (figure 5)

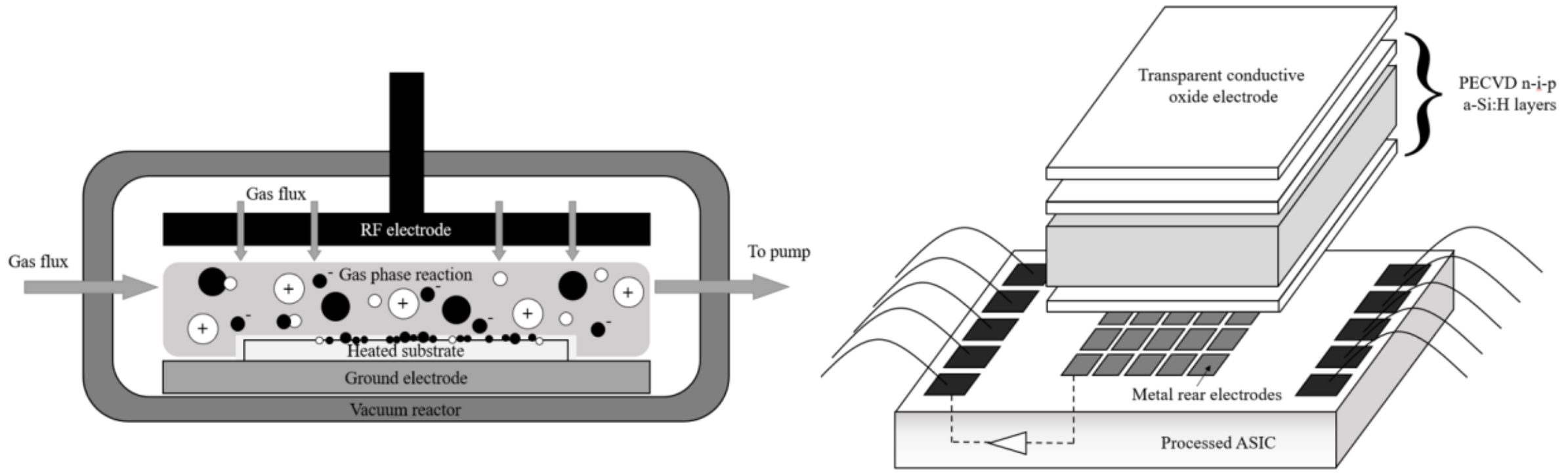

***Figure 5.*** *Typical a-Si:H PECVD schematic (left), and a schematic of a-Si:H deposited directly on ASIC (right).*

Compared to crystalline silicon, a-Si:H has a lower density due to voids that exist in the material. Typical void fraction was found to be around 1 % for device-grade material with an average void diameter of 1 nm, which results in a slightly reduced density value.[33] The band gap of a-Si:H falls in the range of 1.5 to 1.8 eV due to the broad distribution of states, and the ionization energy also varies from 4.8 to 6 eV. [34–36] For amorphous silicon detectors, the active layer is an intrinsic layer. Doping of amorphous silicon introduces additional defects which further reduces the low carrier lifetime and mobility, rendering the doped layers unfit for carrier generation and collection. Thus, detectors with a-Si:H use n-i-p structure where the doped layers provide the internal electric field, and the intrinsic layer generates carriers. The intrinsic layer is typically much thicker than the doped layers (50 ~ 100 $\mu$m vs. 10 nm).

The dangling bonds inside the intrinsic layer of a-Si:H is the result of a broken Si-Si bond and is responsible for its metastability. Unlike the defects in other semiconducting materials, the defect density for a-Si:H is not fixed after the deposition process. This is because external perturbations such as light and heat can result in a new equilibrium. Radiation exposure increases the density of dangling bonds, which leads to the degradation of the material’s electronic properties.[37] This phenomena which is named the Staebler-Wronski effect can be reversed by thermal annealing of the material. The temperature and process time for the annealing process depends on the radiation type as well as the operating temperature of the sensor.

Room temperature annealing is sufficient to recover from proton-induced defects for sensors that are operated at 200 K, whereas the sensors operated at room temperature require temperatures above 400 K.[38] Similar recovery was reported for a-Si:H diodes exposed to high doses of protons. Despeisse et al. irradiated diodes with a 32 $\mu m$ thick intrinsic layer for direct readout of 24 GeV/c proton beam.[39] Due to the high flux of the beam and readout system, the signal resulting from each proton spill was measured. Each spill was approximately $1.6 \times 10^{11}$ protons/cm$^2$. The 2×2 mm sensor showed a decrease to a quarter of its initial proton spill-induced current value after a total fluence of $3.5 \times 10^{15}$ proton/cm$^2$. Sensors annealed at room temperature for 20 hours and under bias recovered to 66 % of the initial proton spill-induced current values, under similar fluences (figure 6). This demonstrates the radiation hardness of a-Si:H detectors, and its potential for beam monitoring and tracking.

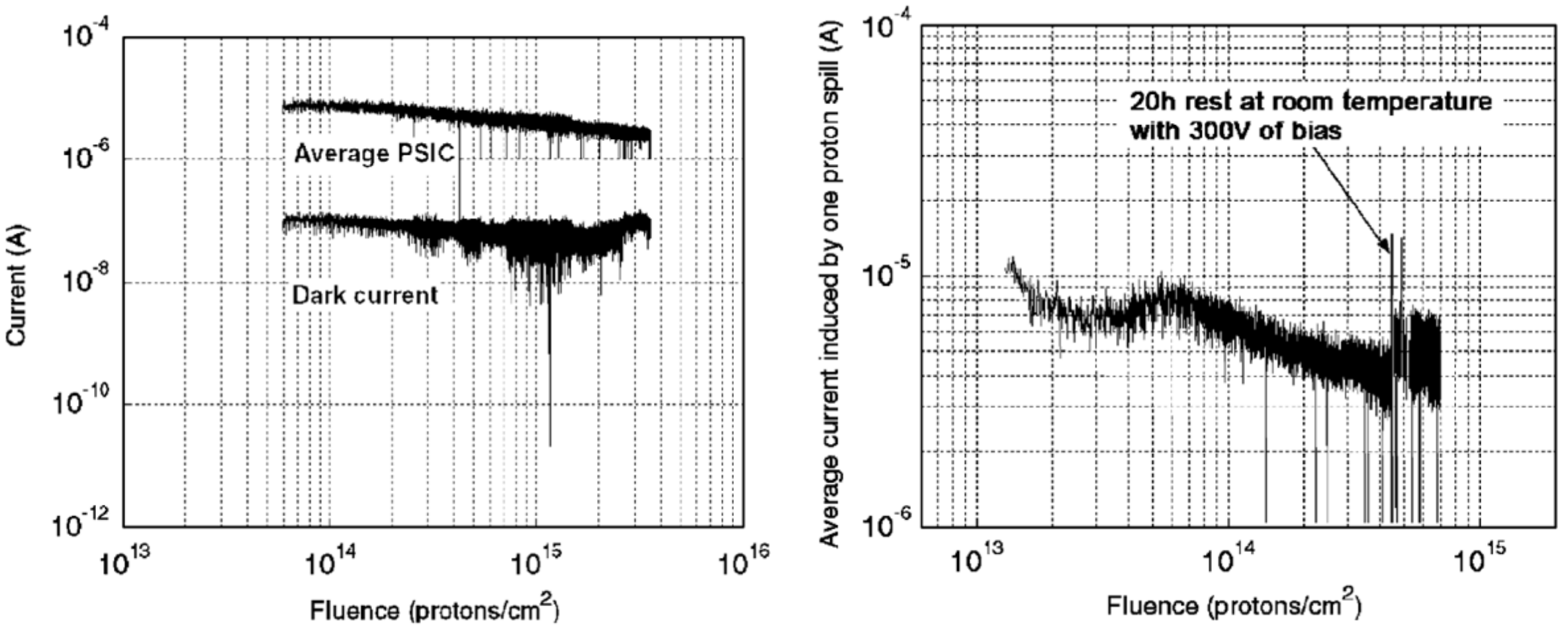

***Figure 6.*** *Proton spill-induced current for a-Si:H diodes in comparison with the dark current. The effects of thermal recovery on the spill-induced current. The device was under bias at room temperature for 20 hours. Reproduced with permission from Ref.* [39]. *Copyright 2005, Elsevier.*

The reverse bias required to achieve the same depletion layer thickness is higher for a-Si:H compared to crystalline silicon, due to the defective nature of the material. This determines how thick the active layer can be, as well as how thick the intrinsic a-Si:H layer should be in the detector. For an ionized defect density of $6 \times 10^{14}\ cm^{-3}$, the required potential to fully deplete a 30 $\mu m$ sensor is about 400 V. For sensors with 50 or 100 $\mu m$ thickness, the bias would be about 1.1 and 4.4 kV, respectively.[40] Increasing reverse bias to create thicker depletion layer is feasible only to a certain extent, as after the threshold the increase in the leakage current begins to drown the signal generated from charged particles. The maximum depletion thickness for a hybrid detector is currently around 40 $\mu m$, thus the detectors are fabricated near that thickness. Thicker sensors than 40 $\mu m$ will result in incomplete depletion, which is undesirable in terms of signal generation due to lower active volume and reduced charge collection.

Despite the relatively low carrier mobility, high reverse bias and reduced detector thickness still allows detector operation with decent carrier drift time.[35, 38, 41–43] A typical electric field strength in a-Si:H diode will be in the range of $10^5$ V/cm, thus the drift times in a 20 $\mu m$ thick diodes are 20 ns for electrons and 1 $\mu s$ for holes. This results in the mean drift length of millimeters and tens of microns for electrons and holes, respectively.

**Direct deposition on ASICs**

The low deposition temperature of PECVD for a-Si:H means that direct deposition on pre-existing ASICs to form a semi-monolithic detector is possible.[35] This type of fabrication method is favorable because it reduces production costs due to the lack for bump-bonding processes, while also improving the signal to noise ratio from the direct interconnection between the sensor element and its electronic counterpart.[40] Vertical integration with the electronics is possible, which preserves the near-unity geometrical fill factor.

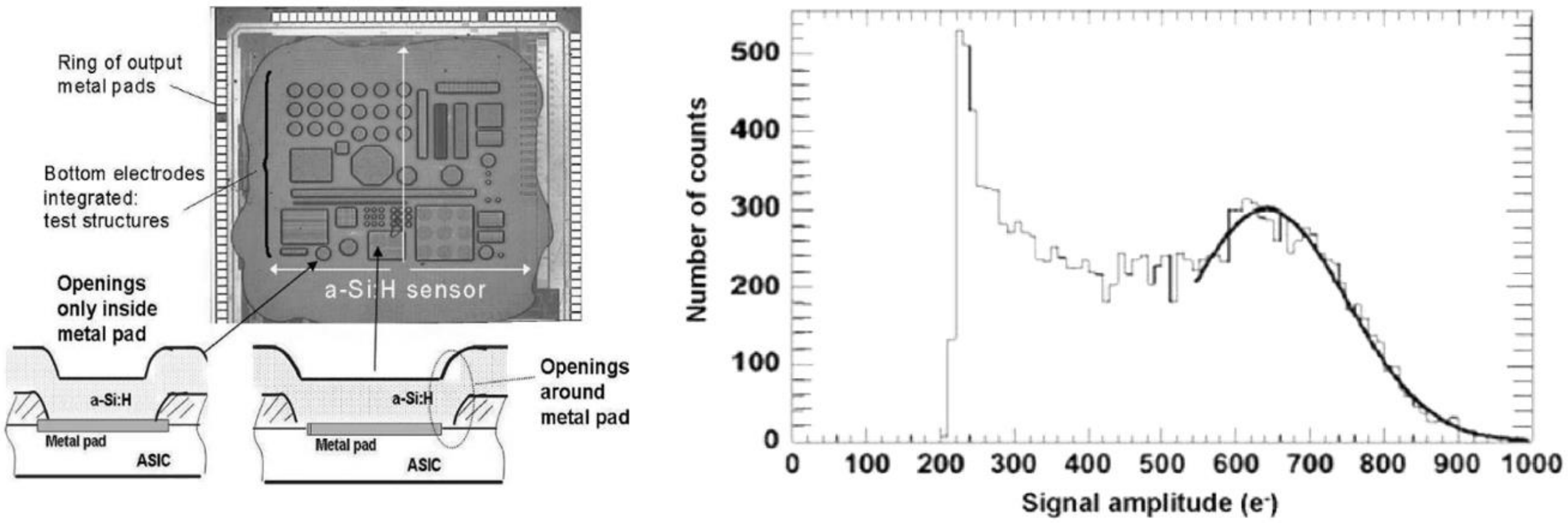

***Figure 7.*** *(Left) Schematic of a thin film on ASIC detector with a 20 μm a-Si:H diode built on top of an integrated circuit. (Right) Spectrum for 5.9 keV X-rays from a* $Fe^{55}$ *source. The sensor was deposited on MACROPAD circuit and operated under a bias of 145 V and threshold of 200* $e^-$*. Reproduced with permission from Ref.* [35]. *Copyright 2008, IEEE.*

Uneven ASIC surfaces from electrode connections, increased leakage currents posed challenges in the first generation thinfilm-ASIC detectors, which were developed around 2007.[35] This led to a decrease in full depletion depth to 20 $\mu m$, about half of detectors using hybrid devices. Overall, a-Si:H is suitable for the fabrication of low ionizing radiation sensor. MIP detection is challenging, but further research on material properties and 3D detectors are underway.[44]

ii. CdTe and CdZnTe

Cadmium telluride (CdTe) and cadmium zinc telluride (CdZnTe) have had some research and applications in radiation detection due to their high atomic number (Z), which offers large photon absorption cross section as well as high energy resolution. The band gap of the compound ranges from 1.44 eV for pure CdTe, to 2.2 eV for pure ZnTe. CdTe is advantageous as it creates a relatively large number of electron-hole pairs (~ 22,000 from $\gamma$-ray with 100 keV). An inherent feature to CdTe is its low hole mobility (~ 1/10 of electron mobility), as well as the limited $\mu\tau$ product for both electrons and holes.

Gädda et al., demonstrated the use of bulk CdTe crystals in pixel detector applications.[45] The sensor components were connected to the electronic components by low temperature indium solder bump bonding, and the resulting device was used to detect the gamma radiation from a $^{137}$Cs source. 0.5 mm thick CdTe layers (Acrorad Ltd. detector grade, > $10^9 \Omega \cdot \mathrm{cm}$) were used as the active layer, and infra-red microscopy was used to probe the crystals of defect density (Fig. 4). The number of defects (grain boundaries, fractures, and Te inclusions) had a direct correlation to the resulting detector's performance. When the number of defects were above the threshold, leakage current became significant, limiting the device's function as a detector.

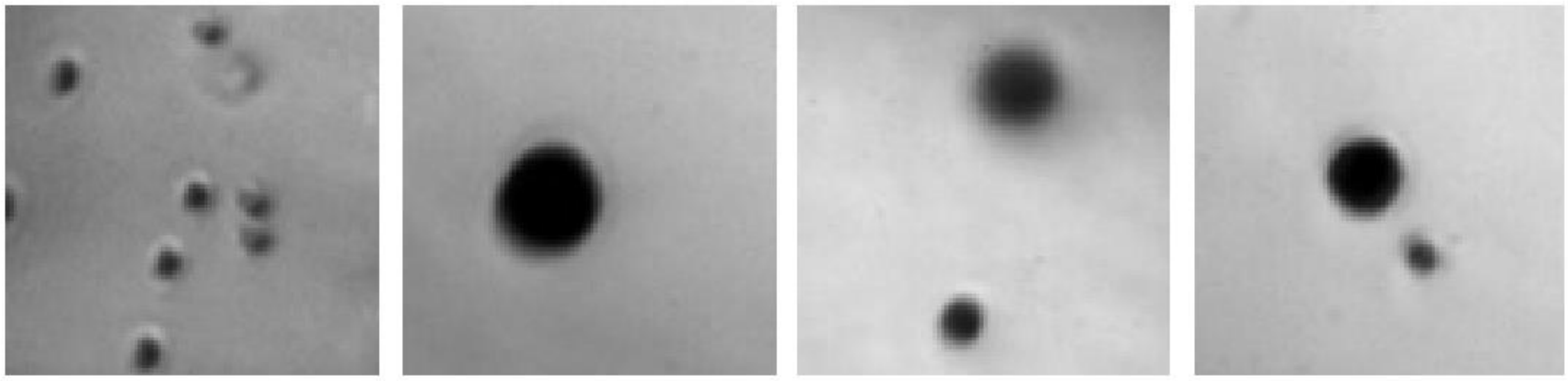

***Figure 8.*** *Typical tellurium precipitates in crystals of CdTe, and CdZnTe. Reproduced with permission from ref*[45]. *© IOP Publishing Ltd and Sissa Medialab. Reproduced by permission of IOP Publishing. All rights reserved*

The 0.5 mm thick hybrid pixel detector was fully depleted at a bias voltage of -100 V and was able to detect 662 keV photons. The energy resolution for this radiation was around 2 %, and due to the selective amplification of electron currents by the CMS readout chip, holes did not contribute to the signal. The sensor showed low quantum efficiency (less than 10 %) for photons with > 600 keV, and the count rate of the electronics had to be lowered at high X-ray beam conditions to prevent the quenching from data buffers, resulting in low count numbers.

The size limitations and defects in CdTe ingots led to Jiang et al., in developing a modified close space sublimation (CSS) method for CdTe films.[46] CSS is a vacuum epitaxial film deposition method commonly used to prepare polycrystalline CdTe/CdS solar cells, thus benefiting from the relatively streamlined process. Single crystalline thick films (~ 200 $\mu m$) with high resistivity of ~ $10^{10}$ $\Omega \cdot cm$ were produced from the modified CSS method, at a temperature of 420 °C. Increasing the Te partial pressure during film growth by using Te-rich sources (Cd:Te = 46:54) resulted in the deposition of CdTe films on Ge substrate with smaller FWHM in XRD measurements, compared to those of films deposited using stoichiometric CdTe source. This was attributed to a) improved interface quality between the CdTe and Ge produced by the Te-rich environment and b) absence of Te precipitates, inclusions, or Te decorated grain boundaries in the IR transmission measurement. Due to the promising properties of the deposited CdTe, a follow-up study was done on its application in particle detection.[47] Alpha particle spectrometry was conducted on a CdTe film sensor, connected to a multichannel analyzer (MCA) which amplifies/records the developed signals. The CdTe film used for this device was deposited using Te-rich (Cd:Te = 54:46) source with a thickness of 200 $\mu m$, and Au electrodes were used to fabricate a device. The particles used were 5.49 MeV uncollimated $^{241}$Am, and the measurements were done in room temperature under vacuum of $2 \times 10^{-2}$ Torr with activity of 200 kBq. The resulting alpha spectra with various bias voltages are depicted in figure.9.

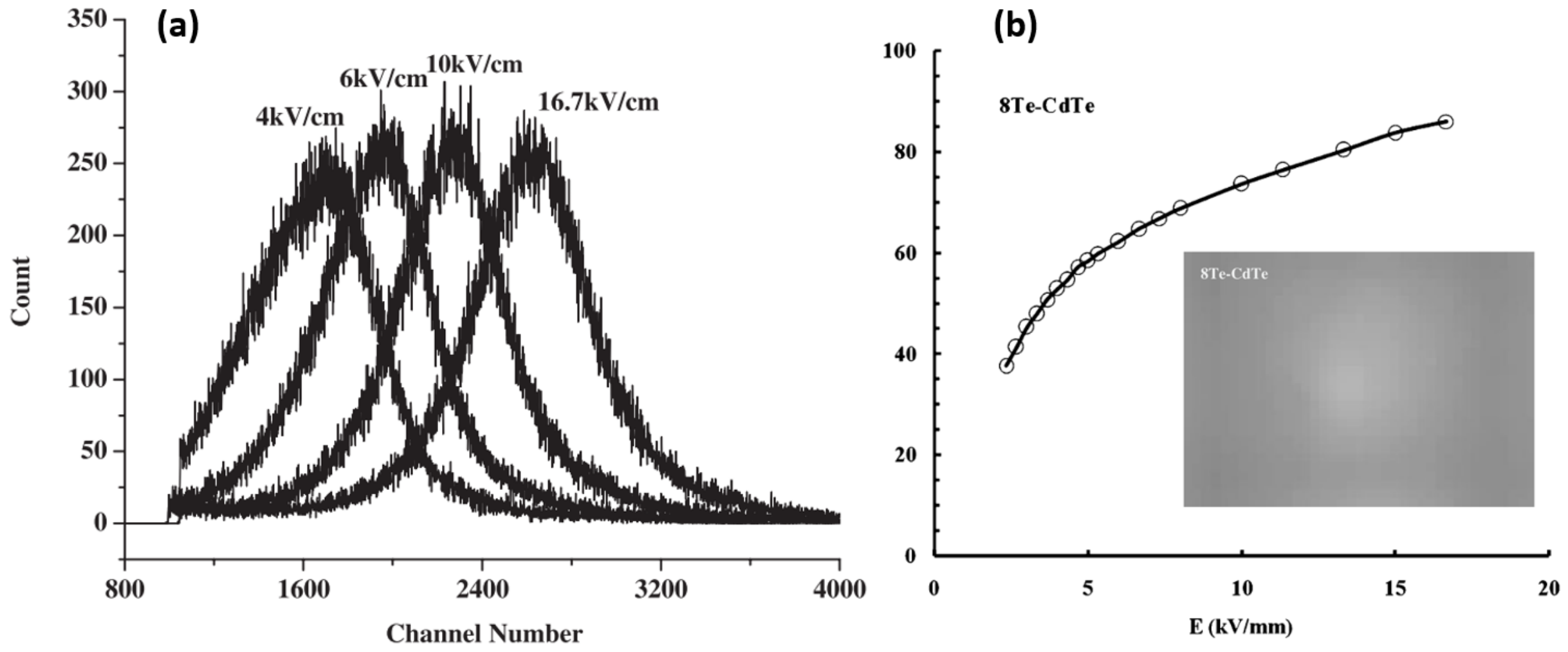

***Figure 9.*** *(a) Alpha particle response for the Au/CdTe/Au device at different driving biases. (b) Hecht plot obtained from the Alpha particle measurements. Inset shows the IR transmission image, indicating the absence of Te inclusions. Reproduced with permission from ref*[47]. *Copyright (2010) The Japan Society of Applied Physics*

The plot shows the peak moving towards higher energy at higher voltages, indicating an increase in efficiency. The Hecht equation was used to calculate the charge collection efficiency (CCE), which also resulted in a $\mu_e\tau_e$ product value of $1.6 \times 10^{-4} cm^2 V^{-1}$ for the Au/CdTe/Au device.

iii. Diamond

Diamond is an insulating material with an indirect band gap of 5.5 eV and almost no intrinsic charge carriers resulting in a high resistivity of $10^{13} - 10^{16}\ \Omega m$. A large displacement energy of 43.3 eV makes diamond exceptionally radiation hard, able to withstand doses of radiation higher than those faced at the large hadron collider (LHC).[48] Due to the lack of intrinsic carriers and large band gap, diamond detectors can be operated at room temperature even after heavy doses of radiation, while maintaining low levels of leakage currents. Due to its high resistivity, diamond detectors operate similar to parallel plate capacitors which have diamond as the insulator. The chemical vapor deposition methods can routine produce wafer-size polycrystalline diamond films.[49, 50] However, with recent advancement in the deposition methods, even monocrystalline diamond films can be deposited over relatively large area (ranging from 1 inch to 3.6 inch in diameter)[51] as monocrystalline films are preferred for detector applications for obvious reasons.

Contact formation

The selective formation of heavily doped, low resistance region is possible through ion implantation and is a common process for typical semiconductor like silicon. Similar process may be used to form contact regions in diamond. However, ion implantation at the temperatures used for silicon result in the destruction of the diamond lattice (graphitization).[52–54] Successful formation of heavily B doped, very low resistive diamond layer has been reported for temperatures around 400 °C.[55] Other interesting methods include depositing low resistive nitrogen incorporated ultrananocrystalline diamond (N-UNCD) layers on diamond providing ohmic contacts.[56]

Polycrystalline growth and the grains

CVD diamond is typically produced via a relatively low-temperatures(450-900$^{o}$C), low-pressure (30-100 Torr) process. Due to the nucleation of carbon atoms on the substrate surface and the following film growth, grain structures are formed with sizes in the range of $\sim 100\ \mu m$ on the growth side, as seen in Fig. 10. During detector fabrication, the film is polished down from the substrate side to a thickness of $\sim 500\ \mu m$. The diamond films could be nanocrystalline or microcrystalline depending on the gas chemistry and growth parameters.[57]

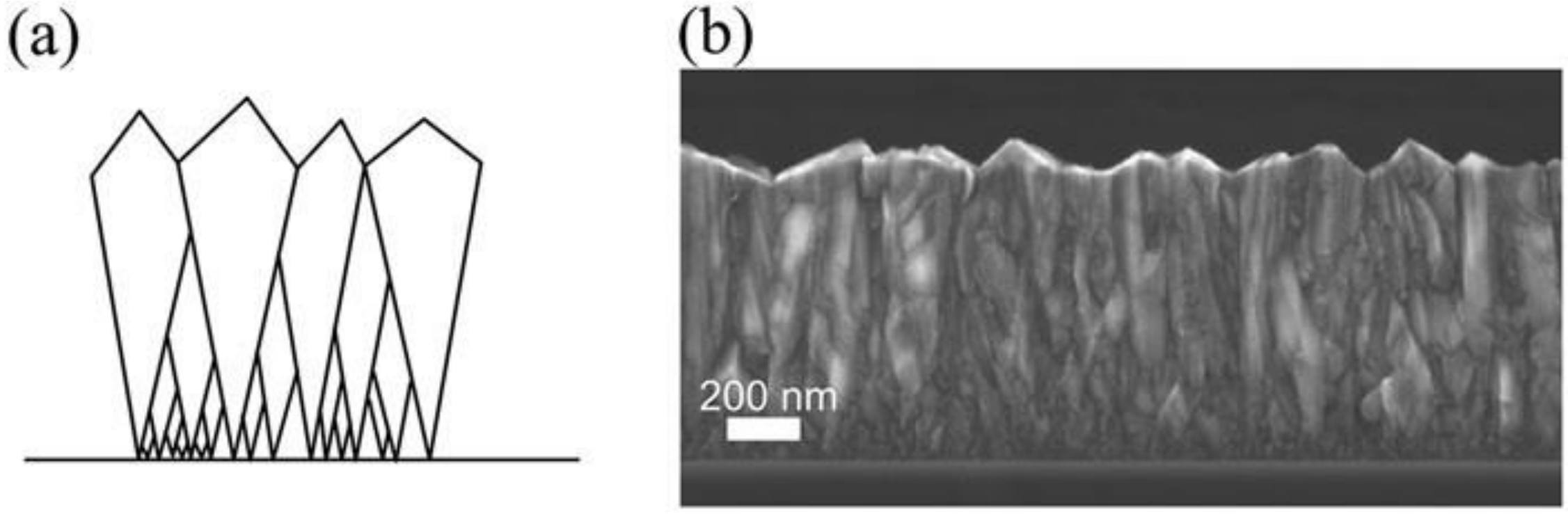

***Figure 10.*** *(a) CVD grown polycrystalline diamond films. (b) SEM cross section image of microcrystalline diamond film. Reproduced with permission from* [58] *under CC BY 4.0.*

Pumping effect of diamond sensors

The response signal of diamond sensors increases during irradiation. This "pumping effect" is the result of long trapping times of the defects, which are typically located deep in the band gap. Thermal emission probability from these transitions is very low, and once filled, they do not trap additional charges. After all traps have been filled, larger signals by a factor of ∼1.5 are observed, due to the increased charge collection distance.

Due to the major advantages of diamond mentioned above, the RD42 collaboration at CERN has been developing polycrystalline CVD diamond for particle detectors in parallel to silicon detectors.[59–66] In 2017, RD42 tested the first diamond 3D pixel detector at the high intensity proton accelerator (HIPA) at the Paul Scherrer institute (PSI).[67] 3D detectors have electrodes located inside the sensing medium which reduces the drift distance to a length much shorter than the medium thickness ($250 - 500\ \mu m$ to $25 - 100\ \mu m$). In these structures, alternating cathodes and anodes are planted in the bulk material, perpendicular to the read-out face. 3D detector design is useful for cases where the mean free path is limited, such as heavily irradiated silicon or polycrystalline CVD diamond. This has the effect of increasing the detector's radiation hardness as it increases the signal strength. Figure 11 (left) shows the efficiency of every cell in the device with a pixel threshold of $1500e$ as a function of xy position. The red box and the blue circle represent the region used to calculate hit efficiency and the location of a single nonfunctioning pixel cell, respectively. Figure 11 (right) is a plot of the hit efficiency in the red square region as a function of time, as the scan rate of incident particle was increased from $7\ kHz/cm^2$ to $7\ MHz/cm^2$ and back down again. The measured overall efficiency was 99.2 %, and the change in the rate of incident particle showed no effect.

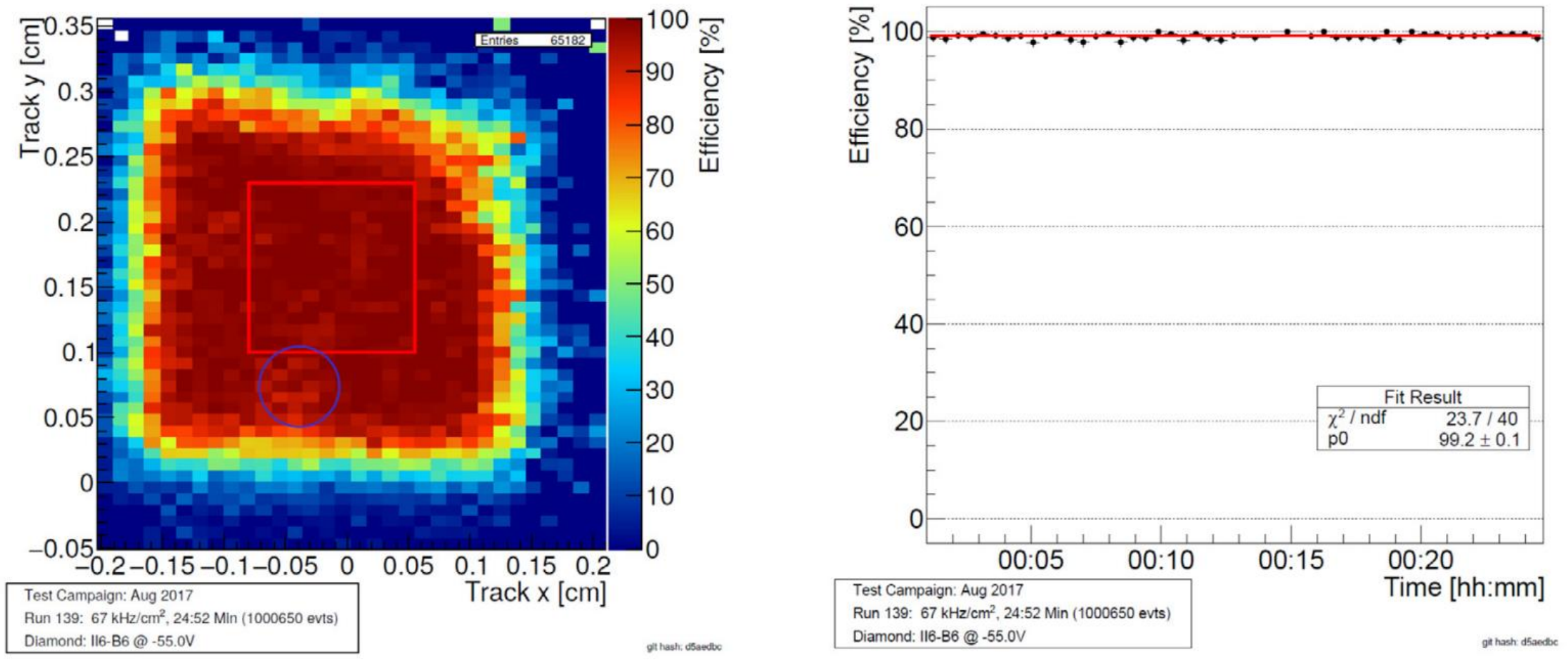

***Figure 11.*** *Hit efficiency of a* $50\ \mu m \times 50\ \mu m$ *cell polycrystalline CVD diamond 3D pixel detector with* $3 \times 2$ *grouped cells, connected to a 1500e threshold CMS pixel electronic. Reproduced with permission from Ref.* [67]. *Copyright 2019, Elsevier.*

Liu et al. have very recently investigated the effect of radiation doses on single crystal diamond (SCD) sensors.[68] SCDs are attractive as a detector medium due to their extreme carrier mobility.[69] The response current of two SCD detectors were continuously measured while the detectors received irradiation of 100 MeV protons with a dose over $10^{17}$ protons/cm$^2$. The SCD wafers were grown via 30 kW DC arc plasma jet CVD on type-Ib (100) single crystal diamond plate. Wafer thickness was approximately $250\ \mu m$. To create an ohmic contact between the metal electrodes and diamond films, Ti-W-Au composite films were deposited by magnetron sputtering followed by thermal annealing. The high kinetic energy from the magnetron sputtering allowed titanium atoms to penetrate the surface of the diamond, and the following annealing process resulted in the formation of titanium carbide. The detectors were operated at an electric field below 2.4 V/$\mu m$, which was the saturated value of the charge collection distance curve. The detectors maintained their functionality throughout the irradiation and showed a decrease in signal intensity to 5 % of the original value. This indicated that the SCD fulfills the requirement for the LHC upgrade. The deterioration of the device performance was affected by the total fluence, as well as the irradiation rate. The current was held at 1 $\mu A$ during the measurement, which corresponds to approximately 80,000 protons per bunch. The decrease in response current was attributed to the formation of defects, which gradually slows down and reaches saturation as seen in Fig. 12.

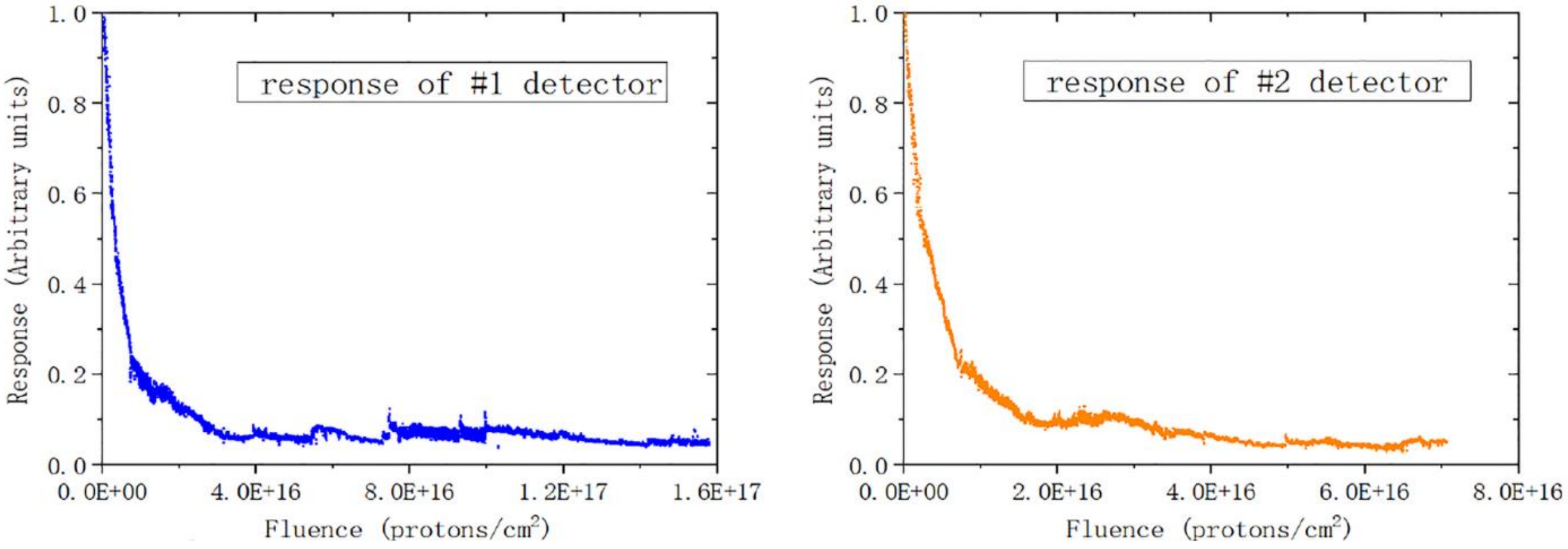

***Figure 12.*** *Normalized current response vs. fluence of two separate diamond detectors. Reproduced with permission from* [68] *under CC BY-NC-ND 4.0.*

iv. GaAs

Gallium arsenide has been used in high speed electronics and photonics devices due to its high carrier mobility ($\mu_e = \sim 10^3\ cm^2V^{-1}s^{-1}, \mu_h = \sim 10^2\ cm^2V^{-1}s^{-1}$).[70] A bandgap of 1.43 eV translates to a lower leakage current compared to silicon under the same reverse biases.[71] The density of GaAs is lower than that of some of the heavier compound semiconductors like CdTe, but higher than that of Si, offering increased efficiency at X-ray energies over 10 keV. However, the high concentration of impurities ($\sim 10^{15}\ cm^{-3}$) in melt-grown bulk gallium typically includes EL2 deep donor defects, which reduces the electron lifetime to the range of several ns, ultimately resulting in low charge collection efficiency. The EL2 defect is a combination of interstitial arsenic atom and an arsenic antisite, and is located 0.65 eV below the conduction band.[72] Typical concentrations are in the order of $10^{16}\ cm^{-3}$. The ionized EL2 defects ($EL2^+$) act as recombination centers with a large cross section of $> 10^{-13}\ cm^2$. Chromium compensation of GaAs (GaAs:Cr) is an effective method to passivate the problematic $EL2^+$ states[73], which can lead to high resistivity ($\sim 10^9\ \Omega cm$) and carrier mobilities while enabling active layers up to 1mm in thickness.[70]

Dachs et al. have recently compared the performance of chromium-compensated gallium arsenide pixel detectors to silicon detectors.[74] The sensors were tested at the CERN SPS facility with 20 GeV/c electrons, muons with 120 ~ 290 GeV/c. 30 Mylar foils 50 $\mu m$ thick were used as radiators, and were positioned approximately 2 m from the detector. The absorption of transition radiation by air was prevented by a pipe filled with helium, located between the radiator and the detector. A Timepix3 chip was bonded to both Si and GaAs:Cr sensor, where the detector consists of a matrix of $256 \times 256$ pixels with a pitch of 55 $\mu m$ with an active area of 1.98 $cm^2$. Both materials were $500\ \mu m$ thick in order to increase detection efficiency of transition radiation. The bias voltage was -450 V for GaAs and -200 V for Si. Projections of the energy-angular distributions are shown in figure 13, where the GaAs:Cr sensor shows better performance in the high-energy photon region. The multiple peaks observed in figure 13 (b) and (c) arise from the interference of the Mylar foils.

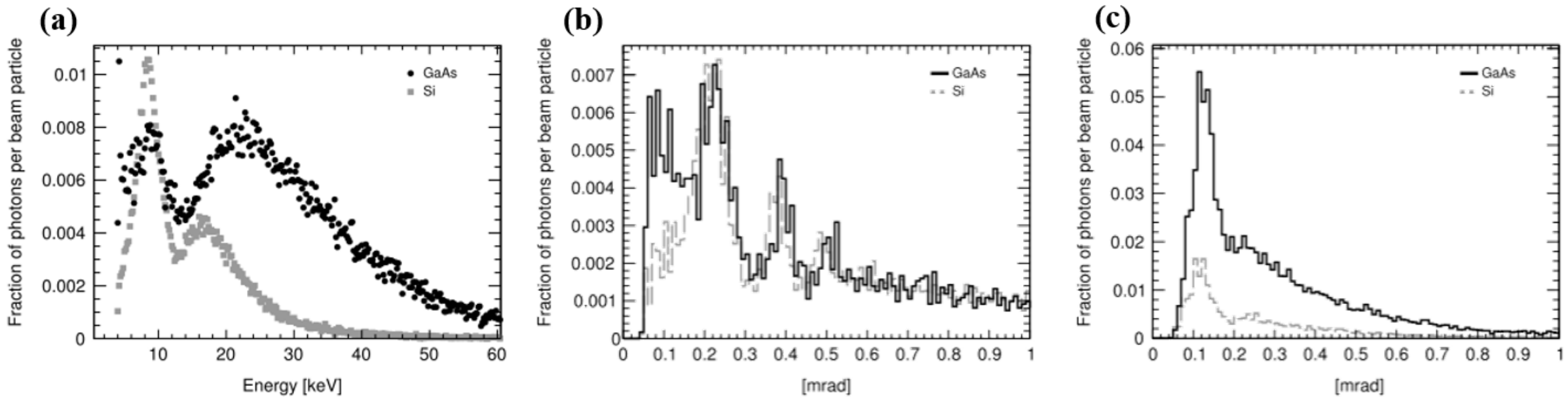

***Figure 13.*** *(a) Energy spectra of Si and GaAs sensors. (b), (c) Angular distribution or transition radiation for Si and GaAs sensors in the photon energy range of under and over 13 keV, respectively. Reproduced with permission from Ref.* [74]. *Copyright 2020, Elsevier.*

v. Perovskite materials

Tsai et. Al. have recently suggested the use of layered 2D perovskite thin film diodes as x-ray detectors.[75] These diodes have a p-i-n junction structure, consisting of a 470 nm thick 2D Ruddlesden-Popper (RP) phase layered perovskite $(BA)_2(MA)_2Pb_3I_{10}(Pb_3)$ which can effectively detect x-ray photons. Due to the high atomic number of the constituting atoms, the absorption coefficient of the perovskite material is between 10 ~ 40 times higher than silicon for x-ray. The ionization energy for the 2D RP was calculated to be 4.46 eV. Due to the large number of carriers generated by the x-ray absorption, a large built-in electric field is created from the quasi-fermi level splitting between the n and p contacts, which facilitate the charge collection. This leads to a unique advantage for the perovskite detector, as the internal electric field is enough to collect the generated charges, removing the need for an applied external potential. The high open-circuit voltage arising from the high density of carrier generation also offers the possibility to use the generated voltage as alternative parameter of detection. For higher energy x-ray detection, thicker layers of 2D RP were required, and thicknesses up to 8 $\mu m$ were deposited with good crystallinity and quality.

Table 3. Summary of the discussed thin-film particle detectors

| | a-Si:H | CdTe | Diamond | GaAs |
|---|---|---|---|---|
| Deposition temperature | ~ 200 °C (PECVD) | ~ 400 °C (CSS) | 450 ~ 950 °C (Nanocrystalline) | 450 ~ 1050 °C[76] (CVD) |
| Carrier drift mobility ($cm^2/(Vs)$) | 1~5 (electron) 0.01 (hole) | 1050 (electron) 100 (hole) | 1800 (electron) 1200 (hole) | 8800 (electron) 320 (hole) |
| Carrier lifetime | ~ $\mu s$ | 0.1-2 $\mu s$ | ~ 1 ns | 1-10 ns |
| Prospects | High depletion voltage, incomplete charge collection, direct deposition on ASICS possible | High atomic number offers large photon absorption cross section, energy resolution. | Radiation hardness, low leakage current, short carrier drift distance | High carrier mobility, requires passivation of deep level defects (bulk grade) |
| Manufacturing capability | Industrial scale already available, due to use in solar industry | Wafer scale deposition for high quality film remains to be achieved | Large scale deposition available (polycrystalline), due to demands in research / gem market | Industrial scale production available |

b. Current challenges of thin film detectors

i. Sensitivity

Strength of the signal depends on how many charge carriers are created in the active layer of the sensor, and how efficient the collection of the charges is. Very thin film detectors may encounter problems due to not enough charge carriers being generated from the traversing radiation. Also, the mobility and lifetime of carriers must be sufficient for the generated carriers to be fully collected through the connected electrodes. Carrier drift time may be reduced by the increasing the electrical bias, but only to a certain extent, as breakdown of the sensor material can occur at very high biases.

ii. Reliability

Durability of the wafer and device is jeopardized when the material becomes very thin. The handling of the substrate requires structural integrity to prevent damage and breaking during processing. This requirement becomes less stringent if the entire process is automated.

Feasibility of using a material for tracking detector purposes is influenced heavily by the currently available material deposition technology. Advances in the technique leads to better material characteristics, while lowering the production cost. Thus, the application of a technology in the field depends heavily on the large-scale reproduction and commercialization of the material production.

iii. Radiation durability

Pixel detectors are positioned in the innermost regions for collider experiments, which exposes them to large fluence and high luminosity, resulting in significant radiation damage. As the effects of radiation damage are well defined for silicon detectors, this section will use silicon as an example of semiconductor radiation damage. Radiation damage in pixel detectors may be categorized into two types: bulk and surface damage.

Bulk damage

Bulk damage refers to the damage resulting from the incident particle-nuclei interaction. Unlike ionization, which is a reversible process, most particle-nuclei interactions are not reversible. The energy required to remove an atom from its lattice varies on the material, with silicon having values of 11-22 eV, germanium of 12-30 eV, and diamond with a value of 35-80 eV.[48] The values are affected by the direction of the crystal structure. The particle type also determines the nature of the interaction between particle-nuclei. Lighter particles such as electrons require much more energy to initiate the nuclei displacement, compared to heavier particles like neutrons and protons. In order to compare damage from different particles and energies, radiation damage is commonly described by the nonionizing energy loss (NIEL). This quantity includes all energy that is deposited into the sensor material which is not used for the ionization process. As a reference, neutrons with an energy of 1 MeV are used. It is expressed in a way that the fluence $\Phi_{phys}$ of a given particle gives rise to the same NIEL as a fluence $\Phi_{eq}$ of 1 MeV neutrons.

-Leakage current

When the defect has energy levels lying in the band gap, they will serve as recombination centers for generated carriers. This leads to a decrease in generation lifetime $\tau_g$ and an increase in the volume generation current $I_{vol}$. Their relationship is described by:

$$\frac{1}{\tau_g} = \frac{1}{\tau_{g,\Phi=0}} + k_\tau \Phi$$

where $k_\tau$ is the lifetime-related damage rate. Post-irradiation leakage current can be annealed at room temperatures or elevated temperatures, where higher temperatures lead to increased rate of annealing.

- Type inversion, increased full depletion voltage

As continuous irradiation begins to increase the concentration of defective doping to the levels of initial donor or acceptor concentration, the effective doping $N_{eff}$ which describes the net doping effect must be considered. The effective doping can be calculated from the full depletion voltage through the following expression:

$$|N_{eff}| = \frac{2\varepsilon_0 \varepsilon_{si} V_{depl}}{ed^2}$$

where d is the thickness of the sensor. The effective doping as a function of fluence is plotted in figure 14.[77] The n-doped silicon changes doping type at a proton fluence of $1.1 \times 10^{13}\ cm^{-2}$, at which point the space charge is almost zero. Further irradiation results in the domination of the acceptor-like defects, and thus the material becomes p-type. This results in the increase of depletion voltage required for sensor operation.

Heavily irradiated silicon has a different electric field inside the sensor from the initial field. Short pulse laser may be used which penetrate into the sensor materials to analyze the pulse evolution and identify the electric field inside.[78]

The above-mentioned bulk damages affect the irradiated sensor operation, where the operation voltage and leakage current are increased. This leads to an increase in heat generation in the sensor, which in turn further increases the leakage current and generated heat. This mandates the use of proper cooling systems in order to prevent thermal runaways. As the sensor becomes increasingly irradiated, type inversion in some areas of the sensor will occur. This increases the required voltage to fully deplete the sensor, in some cases to thousands of volts. Continuously increasing the operation voltage is not always feasible, and thus decisions must be made on whether to operate the sensor at partial depletion which decreases the signal strength. This becomes an important design consideration for detector systems that must endure extended irradiation damage.

- Charge trapping

Defects can act as recombination centers that increase the leakage current but can also act as trapping centers. The trapping centers in the depletion region are usually empty due to the absence of free charge carriers in the region. These traps can trap and hold charge carriers created by the traveling charged particle for longer than the charge collection time, which reduces the strength of the signal. Unlike spectroscopic applications which suffer significant degradation from charge traps, tracking devices for particle physics applications are not as significantly affected. Only at the higher fluences will the reduced charge collection efficiency become an issue (~ 50 % at $10^{15}$ $n_{eq}/cm^2$ for 300 $\mu m$ thick detector).

Surface damage

Surface damage includes all effects from defects in the dielectric layers, as well as the interface between silicon and dielectric. Although the amorphous structure of silicon oxide prevents macroscopic changes from atomic displacements, ionization in the oxide is not fully reversible. The formation of charges in the oxide layer can occur in several ways. First, is by trapping of holes in traps. Holes have relatively small mobility inside the oxide due to the existence of numerous shallow hole traps. Hole conduction occurs through hopping from one trap to the next. Deeper hole traps exist towards the silicon/oxide interface, and when holes reach this area, they may become trapped permanently.[79] It has been suggested that these deep trap sites are interstitial oxygen atoms. Trivalent silicon atoms that have dangling bonds may become a source of positive oxide charge by the removal of the outer electron.[80]

c. Future directions and opportunities

i. Roll-to-roll production of thin films

Recent technological advancements have enabled roll-to-roll (R2R) production methods, where thin films are deposited over flexible substrates that are fed into a reactor using rollers.[81–83] R2R coating technology has the advantage of high throughput compared to conventional batch-production methods due to its ability to continuously coat the substrate throughout the rolling process. Large-area printing and deposition techniques such as roll-to-roll deposition could significantly lower the detector cost resulting in trackers being produced at a fraction of the current cost. The operating pressure for a R2R coating technology can be anywhere from atmospheric to low vacuum, and the technology is compatible with a wide range of already existing deposition techniques, including solution or slurry based blade-coating, chemical vapor deposition, atomic layer deposition, physical vapor deposition, and sputtering.[81] For the purpose of thin film detectors, materials with good crystallinity offer better carrier lifetimes and mobility, which means that the implementation of R2R technology to vacuum deposition methods will be preferable to solution-based deposition methods. In this regard, the R2R technology can drastically increase the efficiency of precursor usage, since less of the costly precursors are wasted in the purge cycles compared to the batch reactors. Sputtering can be used to deposit a range of metals and transparent/conductive oxides for metallization, and the R2R style deposition of such materials has been demonstrated (figure 14).[83] Deposition on flexible substrates will also enable unique detector geometries as those shown in figure 15.

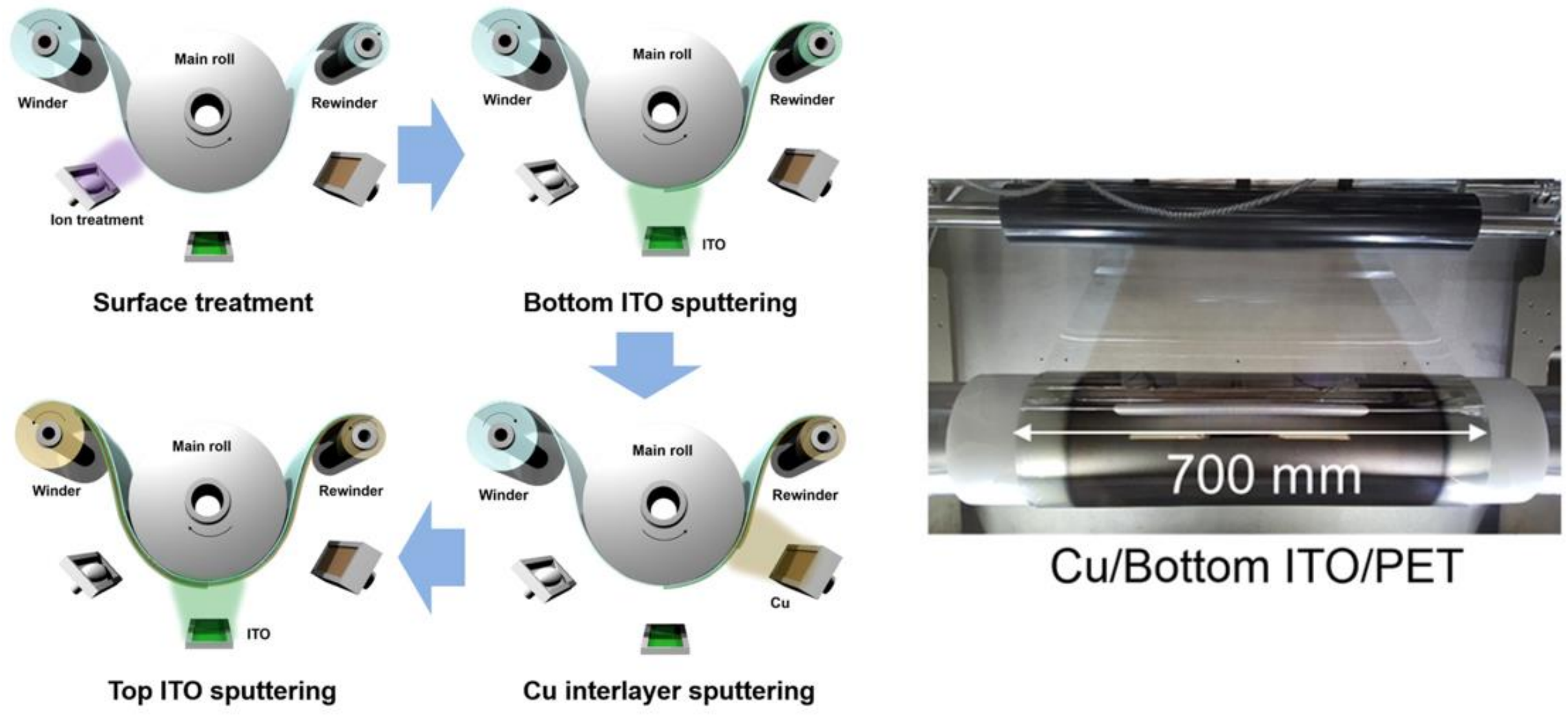

***Figure 14.*** *(Left)A series of coating process using R2R techniques to deposit copper and indium-tin-oxide for transparent thin film heaters. (Right) Picture of a roller with the resulting coated polymer films. Reproduced with permission from* [83] *under CC BY 4.0.*

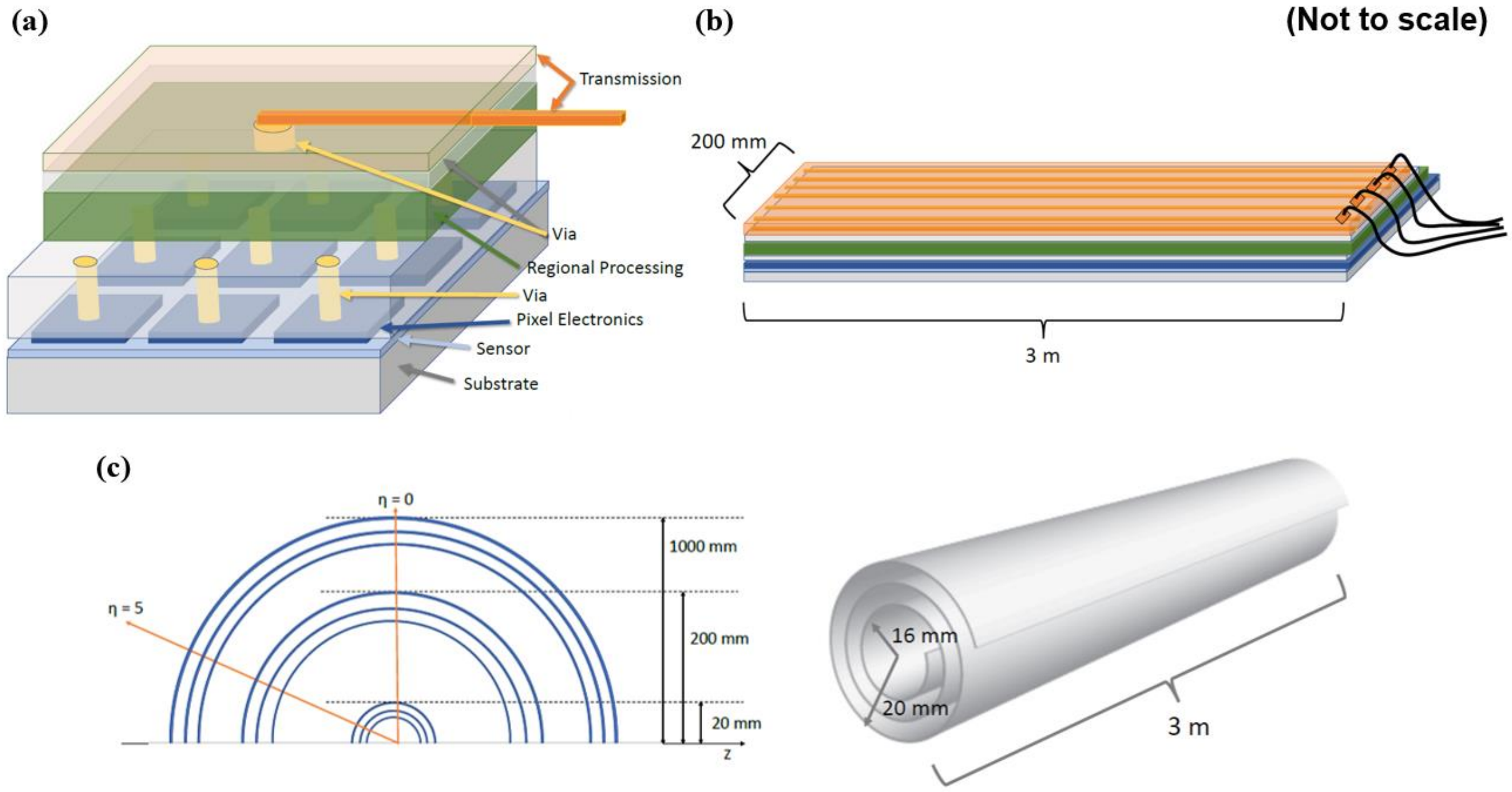

***Figure 15.*** *Monolithic thin film detectors deposited on flexible substrates (a) could be produced in large area sheets (b) which may be rolled into concentric circular tubes surrounding the beam pipe (c).*

ii. Influencing high energy physics experiments

Thin films are a promising candidate because they may build upon techniques that are already developed for other purposes (for example, consumer electronics) that have substantial industrial support, and is likely to have competitive pricing in the future.[84] Thin film technology has the potential to replace parts of the detector, including pixelated trackers and calorimeters. Thin film detectors will also dramatically reduce the amount of dead material in the detectors, which will reduce the amount of unwanted secondary interactions resulting in improved detector performance. Based on the tracking efficiencies expected in the future generation of detectors, track reconstruction efficiencies of 99% will require nuclear interaction lengths that are in the range of ~ 0.01 cm. Such increase in track reconstruction efficiencies will enable a host of new measurements.[85] This will be critical in precision measurements such as track reconstruction, vertexing, and b-tagging.[86]

iii. Medical applications of pixel detectors

Particle physics has made major contribution to the biomedical field, specifically in the area of instrumentation for diagnosis and therapy. The technologies originally developed for particle detection have significantly affected the evolution of medical imaging, most notably in the development of positron emission tomography (PET). Thus, further advances in particle detection technology will likely lead to better imaging performance, or even entirely new imaging techniques and instruments.

## Conclusion

In summary, the advancement of material synthesis and deposition techniques has led to the incorporation of unique and exotic semiconducting/insulating materials into particle detection and tracking applications. By harnessing the properties of these materials, the manufacture of advanced particle detectors which offer high energy resolution, exceptional radiation hardness, operation at room temperature, and monolithic configurations, is possible. Transformative discovery is often driven by advances in instrumentation, and the development of thin film particle detector technology brings humankind closer to the next big discovery in particle physics. Advances in particle detection technology are not only connected to the new discoveries in particle physics, but also provide improved capabilities for medical imaging techniques, such as higher resolution and quicker PET-CT scans.

## Acknowledgement

Funding from Argonne National Laboratory, provided by the Director, Office of Science, of the U.S. Department of Energy under Contract No. DE-AC02-06CH11357. Work performed at the Center for Nanoscale Materials, a U.S. Department of Energy Office of Science User Facility, was supported by the U.S. DOE, Office of Basic Energy Sciences, under Contract No. DE-AC02-06CH11357. Vikas Berry acknowledges financial support from the Office of Naval Research (Contract: N000141812583) and University of Illinois at Chicago.

# References

[1] M. Aaboud, G. Aad, B. Abbott, J. Abdallah, O. Abdinov, B. Abeloos, S. H. Abidi, O. S. AbouZeid, N. L. Abraham, et al., *Performance of the ATLAS Transition Radiation Tracker in Run 1 of the LHC: tracker properties*, *J. Instrum.* **12**, (2017) P05002.

[2] C. A. J. Ammerlaan, R. F. Rumphorst and L. A. C. Koerts, *Particle identification by pulse shape discrimination in the p-i-n type semiconductor detector*, *Nucl. Instruments Methods* **22**, (1963) 189.

[3] R. Eisberg, D. Ingham, M. Makino, R. Cole and C. Waddell, *Semiconductor detector telescopes for measuring proton energies up to 300 MeV — problems and solutions*, *Nucl. Instruments Methods* **101**, (1972) 85.

[4] W. Seibt, K. E. Sundström and P. A. Tove, *Charge collection in silicon detectors for strongly ionizing particles*, *Nucl. Instruments Methods* **113**, (1973) 317.

[5] L. Rossi, P. Fischer, T. Rohe and N. Wermes, *Pixel Detectors*, Springer, (2006).

[6] W. H. Bragg and R. Kleeman, *XXXIX. On the α particles of radium, and their loss of range in passing through various atoms and molecules* , *London, Edinburgh, Dublin Philos. Mag. J. Sci.* **10**, (1905) 318.

[7] C. Avila-Avendano, I. Mejia, R. Garcia-Lozano, L. E. Reyes, S. Rozhdestvenskyy, C. Pham, B. Pradhan, B. E. Gnade and M. A. Quevedo-Lopez, *Integrated Thin-Film Radiation Detectors and In-Pixel Amplification*, *IEEE Trans. Electron Devices* **65**, (2018) 3809.

[8] P. Rinard, *Neutron Interactions with Matter*, in *Passiv. Nondestruct. assay Nucl. Mater.*, (1991).

[9] E. Tupitsyn, P. Bhattacharya, E. Rowe, L. Matei, Y. Cui, V. Buliga, M. Groza, B. Wiggins, A. Burger, et al., *Lithium containing chalcogenide single crystals for neutron detection*, *J. Cryst. Growth* **393**, Elsevier, (2014) 23.

[10] Particle data group, *Review of Particle Physics, 2000-2001*, *Eur. Phys. J. C* **15**, (2000) 1.

[11] J.D.Jackson, *Classical Electrodynamics*, 3rd ed., Wiley, New York, (1998).

[12] C. A. Klein, *Bandgap dependence and related features of radiation ionization energies in semiconductors*, *J. Appl. Phys.* **39**, (1968) 2029.

[13] H. Y. Cho, J. H. Lee, Y. K. Kwon, J. Y. Moon and C. S. Lee, *Measurement of the drift mobilities and the mobility-lifetime products of charge carriers in a CdZnTe crystal by using a transient pulse technique*, *J. Instrum.* **6**, (2011).

[14] G. R. Lynch and O. I. Dahl, *Approximations to multiple Coulomb scattering*, *Nucl. Instruments Methods Phys. Res. Sect. B Beam Interact. with Mater. Atoms* **58**, (1991) 6.

[15] J. H. Laue, *Flip Chip Technologies*, McGraw-Hill, New York, (1996).

[16] O. Ehrmann, G. Engelmann, J. Simon and H. Reichl, *A Bumping Technology for Reduced Pitch*, in *Proc. 2nd Int. TAB Symp.*, San Jose, (1990).

[17] J. Wolf, G. Chmiel and H. Reichl, *Lead/Tin (95/5%) Solder Bumps for Flip Chip Applications Based*

*on Ti:W(N)/Au/Cu Underbump Metallization*, in *Proc. 5th Int. TAB/Advanced Packag. Symp. ITAP*, San Jose, (1993).

[18] A. M. Fiorello, *ATLAS Bump Bonding Process*, in *Pixel 2000 Conf.*, Genoa, Italy, (2000).

[19] P. Fischer, M. Kouda, H. Krüger, M. Lindner, G. Sato, T. Takahashi, S. Watanabe and N. Wermes, *A counting CdTe pixel detector for hard x-ray and γ-ray imaging*, *IEEE Trans. Nucl. Sci.* **48**, (2001) 2401.

[20] W. R. T. Ten Kate and C. L. M. Van der Klauw, *Experimental results on an integrated strip detector with CCD readout*, *Nucl. Instruments Methods Phys. Res. Sect. A Accel. Spectrometers, Detect. Assoc. Equip.* **228**, (1984) 105.

[21] S. Holland, *Fabrication of detectors and transistors on high-resistivity silicon*, *Nucl. Instruments Methods Phys. Res. Sect. A Accel. Spectrometers, Detect. Assoc. Equip.* **275**, (1989) 537.

[22] G. Vanstraelen, I. Debusschere, C. Claeys and G. Declerck, *Fully integrated CMOS pixel detector for high energy particles*, *Nucl. Inst. Methods Phys. Res. A* **275**, (1989) 574.

[23] W. Snoeys, J. Plummer, S. Parker and C. Kenney, *A new integrated pixel detector for high energy physics*, *IEEE Trans. Nucl. Sci.* **39**, (1992) 1263.

[24] I. Perić, *A novel monolithic pixelated particle detector implemented in high-voltage CMOS technology*, *Nucl. Instruments Methods Phys. Res. Sect. A Accel. Spectrometers, Detect. Assoc. Equip.* **582**, (2007) 876.

[25] P. Castelein, J. M. Debono, M. Fendler, C. Louis, F. Marion, L. Mathieu and M. Volpert, *Ultra fine pitch hybridization of large imaging detectors*, *IEEE Nucl. Sci. Symp. Conf. Rec.* **5**, IEEE, (2003) 3518.

[26] D. E. GROOM, N. V. MOKHOV and S. I. STRIGANOV, *MUON STOPPING POWER AND RANGE TABLES 10 MeV–100 TeV*, *At. Data Nucl. Data Tables* **78**, (2001) 183.

[27] D. W. Palmer, *The Semiconductors-Information Web-site*www.semiconductors.co.uk (5 January 2021).

[28] J. Metcalfe, I. Mejia, J. Murphy, M. Quevedo, L. Smith, J. Alvarado, B. Gnade and H. Takai, *Potential of Thin Films for use in Charged Particle Tracking Detectors*, (2014).

[29] A. Beer, R. Willardson and E. Weber, *Semiconductors for Room Temperature Nuclear Detector Applications*, (1995).

[30] D. E. Carlson and C. R. Wronski, *Amorphous silicon solar cell*, *Appl. Phys. Lett.* **28**, (1976) 671.

[31] H. C. Tuan, *Amorpiious Silicon Thin Film Transistor and its Applications to Large-Area Electironics*, *MRS Proc.* **33**, Cambridge University Press, (1984) 247.

[32] A. E. Owen, P. G. Le Comber, W. E. Spear and J. Hajto, *Memory switching in amorphous silicon devices*, *J. Non. Cryst. Solids* **59**–**60**, (1983) 1273.

[33] M. J. Van Den Boogaard, S. J. Jones, Y. Chen, D. L. Williamson, R. A. Hakvoort, A. Van Veen, A. C. Van Der Steege, W. M. A. Bik, W. G. J. H. M. Van Sark, et al., *The Influence of the Void Structure on Deuterium Diffusion in a-Si:H*, *MRS Proc.* **258**, (1992) 407.

[34] V. Perez-Mendez, S. N. Kaplan, G. Cho, I. Fujieda, S. Qureshi and W. Ward, *Hydrogenated*

*Amorphous Silicon Pixel Detectors for Minimum Ionizing Particles*, *Nucl. Instruments Methods Phys. Res.*, (1988).

[35] M. Despeisse, G. Anelli, P. Jarron, J. Kaplon, D. Moraes, A. Nardulli, F. Powolny and N. Wyrsch, *Hydrogenated Amorphous Silicon Sensor Deposited on Integrated Circuit for Radiation Detection*, *IEEE Trans. Nucl. Sci.* **55**, (2008) 802.

[36] S. Najar, B. Equer and N. Lakhoua, *Electronic transport analysis by electron-beam-induced current at variable energy of thin-film amorphous semiconductors*, *J. Appl. Phys.* **69**, (1991) 3975.

[37] D. L. Staebler and C. R. Wronski, *Reversible conductivity changes in discharge-produced amorphous Si*, *Appl. Phys. Lett.* **31**, (1977) 292.

[38] N. Kishimoto, H. Amekura, K. Kono and C. G. Lee, *Stable photoconductivity in metastable a-Si:H under high-energy proton irradiation*, *J. Non. Cryst. Solids* **227**–**230**, (1998) 238.

[39] M. Despeisse, P. Jarron, K. M. Johansen, D. Moraes, A. Shah and N. Wyrsch, *Preliminary radiation tests of 32µm thick hydrogenated amorphous silicon films*, *Nucl. Instruments Methods Phys. Res. Sect. A Accel. Spectrometers, Detect. Assoc. Equip.* **552**, (2005) 88.

[40] K. Iniewski, *Semiconductor Radiation Detection Systems*, CRC Press, Boca Raton, (2010).

[41] C. Hordequin, A. Brambilla, P. Bergonzo and F. Foulon, *Nuclear radiation detectors using thick amorphous-silicon MIS devices*, *Nucl. Instruments Methods Phys. Res. Sect. A Accel. Spectrometers, Detect. Assoc. Equip.* **456**, (2001) 284.

[42] J. R. Srour, G. J. Vendura, D. H. Lo, C. M. C. Toporow, M. Dooley, R. P. Nakano and E. E. King, *Damage mechanisms in radiation-tolerant amorphous silicon solar cells*, *IEEE Trans. Nucl. Sci.* **45**, (1998) 2624.

[43] R. A. Street, R. B. Apte, T. Granberg, P. Mei, S. E. Ready, K. S. Shah and R. L. Weisfield, *High performance amorphous silicon image sensor arrays*, *J. Non. Cryst. Solids* **227**–**230**, (1998) 1306.

[44] M. Menichelli, M. Boscardin, M. Crivellari, J. Davis, S. Dunand, L. Fanò, F. Moscatelli, M. Movileanu-Ionica, M. Petasecca, et al., *3D Detectors on Hydrogenated Amorphous Silicon for particle tracking in high radiation environment*, *J. Phys. Conf. Ser.* **1561**, (2020) 012016.

[45] A. Gädda, A. Winkler, J. Ott, J. Härkönen, A. Karadzhinova-Ferrer, P. Koponen, P. Luukka, J. Tikkanen and S. Vähänen, *Advanced processing of CdTe pixel radiation detectors*, *J. Instrum.* **12**, (2017).

[46] Q. Jiang, A. W. Brinkman, B. J. Cantwell, J. T. Mullins, F. Dierre, A. Basu, P. Veeramani and P. Sellin, *Growth of thick epitaxial CdTe films by close space sublimation*, *J. Electron. Mater.* **38**, (2009) 1548.

[47] Q. Jiang, A. W. Brinkman, P. Veeramani and P. J. Sellin, *Epitaxial growth of high-resistivity CdTe thick films grown using a modified close space sublimation method*, *Jpn. J. Appl. Phys.* **49**, (2010).

[48] J. Koike, D. M. Parkin and T. E. Mitchell, *Displacement threshold energy for type IIa diamond*, *Appl. Phys. Lett.* **60**, (1992) 1450.

[49] H. Kagan, *Recent advances in diamond detector development*, *Nucl. Instruments Methods Phys. Res. Sect. A Accel. Spectrometers, Detect. Assoc. Equip.* **541**, (2005) 221.

[50] S. Koizumi, H. Umezawa, J. Pernot and M. Suzuki, *Diamond Wafer Technologies for Semiconductor Device Applications*, in *Power Electron. Device Appl. Diam. Semicond.*, Woodhead Publishing, (2018).

[51] Q. Wei, F. Lin, R. Wang, X. Zhang, G. Chen, J. Hussain, D. He, Z. Zhang, G. Niu, et al., *Heteroepitaxy growth of single crystal diamond on Ir/Pd/Al2O3 (11–20) substrate*, *Mater. Lett.* **303**, Elsevier B.V., (2021) 130483.

[52] C. Uzan-Saguy, C. Cytermann, R. Brener, V. Richter, M. Shaanan and R. Kalish, *Damage threshold for ion-beam induced graphitization of diamond*, *Appl. Phys. Lett.* **67**, (1995) 1194.

[53] F. Fontaine, E. Gheeraert and A. Deneuville, *Conduction mechanisms in boron implanted diamond films*, *Diam. Relat. Mater.* **5**, (1996) 752.

[54] R. Kalish, A. Reznik, S. Prawer, D. Saada and J. Adler, *Ion-implantation-induced defects in diamond and their annealing: Experiment and simulation*, *Phys. Status Solidi Appl. Res.* **174**, (1999) 83.

[55] N. Tsubouchi, M. Ogura, A. Chayahara and H. Okushi, *Formation of a heavily B doped diamond layer using an ion implantation technique*, *Diam. Relat. Mater.* **17**, (2008) 498.

[56] M. Zou, M. Gaowei, T. Zhou, A. V. Sumant, C. Jaye, D. A. Fisher, J. Bohon, J. Smedley and E. M. Muller, *An all-diamond X-ray position and flux monitor using nitrogen-incorporated ultra-nanocrystalline diamond contacts*, *J. Synchrotron Radiat.* **25**, International Union of Crystallography, (2018) 1060.

[57] J. E. Butler and A. V. Sumant, *The CVD of nanodiamond materials*, *Chem. Vap. Depos.* **14**, (2008) 145.

[58] V. V. Mitic, H. J. Fecht, M. Mohr, G. Lazovic and L. Kocic, *Exploring fractality of microcrystalline diamond films*, *AIP Adv.* **8**, (2018).

[59] C. Bauer, I. Baumann, C. Colledani, J. Conway, P. Delpierre, F. Djama, W. Dulinski, A. Fallou, K. K. Gan, et al., *Recent results from the RD42 Diamond Detector Collaboration*, *Nucl. Instruments Methods Phys. Res. Sect. A Accel. Spectrometers, Detect. Assoc. Equip.* **383**, (1996) 64.

[60] W. Adam, C. Bauer, E. Berdermann, P. Bergonzo, F. Bogani, E. Borchi, A. Brambilla, M. Bruzzi, C. Colledani, et al., *First bump-bonded pixel detectors on CVD diamond*, *Nucl. Instruments Methods Phys. Res. Sect. A Accel. Spectrometers, Detect. Assoc. Equip.* **436**, (1999) 326.

[61] W. Adam, C. Bauer, E. Berdermann, P. Bergonzo, F. Bogani, E. Borchi, A. Brambilla, M. Bruzzi, C. Colledani, et al., *Review of the development of diamond radiation sensors*, *Nucl. Instruments Methods Phys. Res. Sect. A Accel. Spectrometers, Detect. Assoc. Equip.* **434**, (1999) 131.

[62] D. Meier, W. Adam, C. Bauer, E. Berdermann, P. Bergonzo, F. Bogani, E. Borchi, M. Bruzzi, C. Colledani, et al., *Proton irradiation of CVD diamond detectors for high-luminosity experiments at the LHC*, *Nucl. Instruments Methods Phys. Res. Sect. A Accel. Spectrometers, Detect. Assoc. Equip.* **426**, (1999) 173.

[63] W. Adam, E. Berdermann, P. Bergonzo, W. De Boer, F. Bogani, E. Borchi, A. Brambilla, M. Bruzzi, C. Colledani, et al., *The development of diamond tracking detectors for the LHC*, *Nucl. Instruments Methods Phys. Res. Sect. A Accel. Spectrometers, Detect. Assoc. Equip.* **514**, (2003) 79.

[64] W. Adam, B. Bellini, E. Berdermann, P. Bergonzo, W. De Boer, F. Bogani, E. Borchi, A. Brambilla,

M. Bruzzi, et al., *Status of the R&D activity on diamond particle detectors*, *Nucl. Instruments Methods Phys. Res. Sect. A Accel. Spectrometers, Detect. Assoc. Equip.* **511**, (2003) 124.

[65] M. Alviggi, V. Canale, S. Patricelli, L. Merola, A. Aloisio and G. Chiefari, *The ATLAS experiment at the CERN Large Hadron Collider*, *J. Instrum.* **3**, (2008) 8003.

[66] D. Asner, M. Barbero, V. Bellini, V. Belyaev, J. M. Brom, M. Bruzzi, D. Chren, V. Cindro, G. Claus, et al., *Diamond pixel modules*, *Nucl. Instruments Methods Phys. Res. Sect. A Accel. Spectrometers, Detect. Assoc. Equip.* **636**, (2011) 125.

[67] H. Kagan, A. Alexopoulos, M. Artuso, F. Bachmair, L. Bäni, M. Bartosik, J. Beacham, H. Beck, V. Bellini, et al., *Diamond detector technology, status and perspectives*, *Nucl. Instruments Methods Phys. Res. Sect. A Accel. Spectrometers, Detect. Assoc. Equip.* **924**, (2019) 297.

[68] Y. hang Liu, C. wei Loh, J. liang Zhang, F. liang Wu, M. Qi, L. fu Hei, F. xiu Lv, Y. long Lv, T. Ge, et al., *Proton irradiation tests of single crystal diamond detector at CIAE*, *Nucl. Mater. Energy* **22**, Elsevier, (2020) 100735.

[69] J. Isberg, J. Hammersberg, E. Johansson, T. Wikström, D. J. Twitchen, A. J. Whitehead, S. E. Coe and G. A. Scarsbrook, *High carrier mobility in single-crystal plasma-deposited diamond*, *Science (80-. ).* **297**, (2002) 1670.

[70] A. V. Tyazhev, D. L. Budnitsky, O. B. Koretskay, V. A. Novikov, L. S. Okaevich, A. I. Potapov, O. P. Tolbanov and A. P. Vorobiev, *GaAs radiation imaging detectors with an active layer thickness up to 1 mm*, *Nucl. Instruments Methods Phys. Res. Sect. A Accel. Spectrometers, Detect. Assoc. Equip.* **509**, (2003) 34.

[71] M. Rogalla, *Systematic Investigation of GaAs Radiation Detectors for HEP Experiments*, (1997).

[72] B. K. Meyer, D. M. Hofmann, J. R. Niklas and J. M. Spaeth, *Arsenic antisite defect AsGa and EL2 in GaAs*, *Phys. Rev. B* **36**, (1987) 1332.

[73] M. C. Veale, S. J. Bell, D. D. Duarte, M. J. French, A. Schneider, P. Seller, M. D. Wilson, A. D. Lozinskaya, V. A. Novikov, et al., *Chromium compensated gallium arsenide detectors for X-ray and γ-ray spectroscopic imaging*, *Nucl. Instruments Methods Phys. Res. Sect. A Accel. Spectrometers, Detect. Assoc. Equip.* **752**, Elsevier, (2014) 6.

[74] F. Dachs, J. Alozy, N. Belyaev, B. L. Bergmann, M. van Beuzekom, T. R. V. Billoud, P. Burian, P. Broulim, M. Campbell, et al., *Transition radiation measurements with a Si and a GaAs pixel sensor on a Timepix3 chip*, *Nucl. Instruments Methods Phys. Res. Sect. A Accel. Spectrometers, Detect. Assoc. Equip.* **958**, Elsevier Ltd, (2020) 162037.

[75] H. Tsai, F. Liu, S. Shrestha, K. Fernando, S. Tretiak, B. Scott, D. T. Vo, J. Strzalka and W. Nie, *A sensitive and robust thin-film x-ray detector using 2D layered perovskite diodes*, *Sci. Adv.* **6**, (2020) 1.

[76] D. H. Reep and S. K. Ghandhi, *Deposition of GaAs Epitaxial Layers by Organometallic CVD: Temperature and Orientation Dependence*, *J. Electrochem. Soc.* **130**, (1983) 675.

[77] D. Pitzl, N. Cartiglia, B. Hubbard, D. Hutchinson, J. Leslie, K. O'Shaughnessy, W. Rowe, H. F. W. Sadrozinski, A. Seiden, et al., *Type inversion in silicon detectors*, *Nucl. Inst. Methods Phys. Res. A* **311**, (1992) 98.

[78] V. Eremin and Z. Li, *Determination of the Fermi level position for neutron irradiated high resistivity silicon detectors and materials using the transient charge technique (TChT)*, *IEEE Trans. Nucl. Sci.* **41**, (1994) 1907.

[79] G. Harbeke and M. J. Schulz, *Semiconductor Silicon*, 1st ed., Springer Berlin Heidelberg, (1989).

[80] C. T. Sah, *Origin of Interface States and Oxide Charges Generated by Ionizing Radiation*, *IEEE Trans. Nucl. Sci.* **23**, (1976) 1563.

[81] J. Park, K. Shin and C. Lee, *Roll-to-Roll Coating Technology and Its Applications: A Review*, *Int. J. Precis. Eng. Manuf.* **17**, (2016) 537.

[82] P. S. Maydannik, T. O. Kääriäinen, K. Lahtinen, D. C. Cameron, M. Söderlund, P. Soininen, P. Johansson, J. Kuusipalo, L. Moro, et al., *Roll-to-roll atomic layer deposition process for flexible electronics encapsulation applications*, *J. Vac. Sci. Technol. A Vacuum, Surfaces, Film.* **32**, (2014) 051603.

[83] S. H. Park, S. J. M. Lee, E. H. Ko, T. H. Kim, Y. C. Nah, S. J. M. Lee, J. H. Lee and H. K. Kim, *Roll-to-Roll sputtered ITO/Cu/ITO multilayer electrode for flexible, transparent thin film heaters and electrochromic applications*, *Sci. Rep.* **6**, Nature Publishing Group, (2016) 1.

[84] G. Crawford, *DOE Basic Research Needs for High Energy Physics Detector Research & Development*, (2020).

[85] J. Metcalfe, I. Mejia, J. Murphy, M. Quevedo, L. Smith, J. Alvarado, B. Gnade and H. Takai, *Potential of Thin Films for use in Charged Particle Tracking Detectors*, (2014).

[86] M. R. Hoeferkamp, S. Seidel, S. Kim, J. Metcalfe, A. Sumant, H. Kagan, W. Trischuk, M. Boscardin, G.-F. D. Betta, et al., *Novel Sensors for Particle Tracking: a Contribution to the Snowmass Community Planning Exercise of 2021*, (2022) 1.